\documentclass{IEEEtran}
\usepackage[boxruled]{algorithm2e}
\usepackage{enumerate}
\usepackage{cite}
\usepackage{graphicx}
\usepackage{psfrag}
\usepackage{subfigure}
\usepackage{url}
\usepackage{amsmath}
\usepackage{amsthm}
\usepackage{array}
\usepackage{amssymb}
\usepackage{amsfonts}
\usepackage{float}

\newtheorem*{conjecture*}{Conjecture}
\newtheorem{lemma}{Lemma}

\newtheorem{corollary}{Corollary}
\newtheorem{remark}{Remark}
\newtheorem{definition}{Definition}
\newtheorem{theorem}{Theorem}
\newtheorem{example}{Example}

\title{A class of index coding problems with rate $\frac{1}{3}$}
\begin{document}
\author{
\IEEEauthorblockN{Prasad Krishnan and V. Lalitha,\\}
\IEEEauthorblockA{Signal Processing and Communications Research Centre,\\
International Institute of Information Technology, Hyderabad.\\
Email: \{prasad.krishnan,lalitha.v\}@iiit.ac.in\\}
}
\date{\today}
\maketitle
\thispagestyle{empty}	
\pagestyle{empty}
\begin{abstract}
An index coding problem with $n$ messages has symmetric rate $R$ if all $n$ messages can be conveyed at rate $R$. In a recent work, a class of index coding problems for which symmetric rate $\frac{1}{3}$ is achievable was characterised using special properties of the side-information available at the receivers. In this paper, we show a larger class of index coding problems (which includes the previous class of problems) for which symmetric rate $\frac{1}{3}$ is achievable. In the process, we also obtain a stricter necessary condition for rate $\frac{1}{3}$ feasibility than what is known in literature. 
\end{abstract}
\section{Introduction}
\label{sec1}
Index Coding, introduced in \cite{BiK}, considers the problem of efficiently broadcasting a number of messages available at a source, to receivers that already possess some prior knowledge of the messages. In the index coding framework, the source is allowed to encode the messages (and thereby use the channel efficiently) while satisfying the receiver demands. The general class of \textit{groupcast} index coding problems consists of $n$ messages generated at a source, where each message is demanded by at least one receiver. 
Index coding problems where each receiver demands a unique message are called \textit{single unicast index coding} problems and are the most widely studied class.

Although index coding continues to be open in general, several researchers have made inroads into characterising the \textit{rate of index coding}\footnote{roughly, the ratio of information conveyed by every message to the number of times the broadcast channel is used} and presenting achievable schemes. The landmark paper \cite{BBJK} famously connected the scalar linear index coding problem to finding a quantity called \textit{minrank} associated with the \textit{side-information graph} related to the given single unicast index coding problem. Upper and lower bounds on the rate for single unicast index coding have been presented via graph theoretic ideas like clique cover, chromatic number \cite{BiK,BBJK}, local chromatic number \cite{SDL}, fractional clique covering and hyperclique covering \cite{BKL1,BKL2, ArK}, and recently, the `generalized interlinked cycle cover' \cite{TOJ}. Many of these papers naturally lead to constructions of (scalar and vector) linear index codes. Linear codes however are not always found to be optimal \cite{LuS}. Random coding approaches to index coding were studied in \cite{ABKSW}. Bounds on the rate of groupcast index coding were presented in \cite{TDN}.

Interference alignment, well known as a powerful tool to study degrees of freedom in wireless interference networks, was employed to the linear index coding problem in \cite{MCJ1}, by modelling the unavailable side-information as interference. The idea is to assign precoding matrices to the message vectors such that all receivers can decode even in the presence of interference (thereby requiring some degree of linear independence between the precoding matrices, and hence reducing the rate). However, at the same time  the interference at the receivers must be as `aligned' as possible (in order to reduce the amount of linear independence required, i.e. to increase the rate).  This technique was further explored in \cite{SDL,MCJ2,Jaf} and several classes of index coding instances with certain feasible rates were identified based on the properties of the interference seen by the receivers.

This work builds primarily upon the results in \cite{Jaf}. In \cite{Jaf}, a necessary and sufficient condition for the feasibility of \textit{rate half}\footnote{for every two $\mathbb F$ (some finite field) symbols transmitted through the channel, one $\mathbb F$ symbol of each message is conveyed.} in a groupcast index coding problem was established based on the properties satisfied by two graphs obtained from the interference structure of the problem, called the \textit{conflict graph} and the \textit{alignment graph}. Also, a sufficient (but not necessary) condition on the structure of these graphs was given for rate $\frac{1}{3}$ feasibility. Prior work \cite{BBJK} also gives a necessary (but not sufficient) condition for rate $\frac{1}{3}$ feasibility based on the interference structure of the problem. The relevant details of the prior work are discussed elaborately in the forthcoming sections of this paper.

\subsection{Contributions}
\begin{itemize}
\item Firstly, we revise the definition of the conflict graph given in \cite{Jaf}. The conflict graph definition in \cite{Jaf} does not capture the interference structure of the index coding problem sufficiently. In order to rectify this, we define the \textit{conflict hypergraph}, which is shown to capture the interference structure sufficiently. (Subsection \ref{conflicthypergraphsubsec})
\item Our main result shows that rate $\frac{1}{3}$ is achievable in a given index coding problem under certain conditions on the topology of the alignment graph and conflict hypergraph of the index coding problem. The sufficient condition which we present for rate $\frac{1}{3}$ feasibility is looser than the sufficient condition shown in \cite{Jaf}. Therefore, this class of index coding problems is bigger than the previously known class of rate $\frac{1}{3}$ feasible problems. The achievability of rate $\frac{1}{3}$ in such problems is shown by presenting a construction of an index code by random generation of precoding vectors over a large field. (Subsection \ref{rateonethirdfeasibleproblems})
\item Towards obtaining our main result, we also obtain a necessary condition for rate $\frac{1}{3}$ feasibility, which is stricter than the prior condition from \cite{BBJK} (Subsection \ref{necessaryconditionrateonethirds}). Our feasibility conditions can thus be seen as being midway between those of \cite{Jaf} and \cite{BBJK}. 
\end{itemize}
\textit{Notations:} Throughout the paper, we use the following notations. Let $[1:m]$ denote $\{1,2,...,m\}$. For a set of vectors $A$, $sp(A)$ denotes their span. For a vector space $V$, $dim(V)$ denotes its dimension. An arbitrary finite field is denoted by $\mathbb F$. A vector from the $m$-dimensional vector space ${\mathbb F}^m$ is said to be picked \textit{at random} if it is selected according to the uniform distribution on ${\mathbb F}^m$.
\section{Review of Index Coding}
\label{sec2}
Formally, the index coding problem (over some field $\mathbb{F}$) consists of a broadcast channel which can carry symbols from $\mathbb F$, along with the following.
\begin{itemize}
\item A set of $T$ receivers
\item A source which has messages ${\cal W}=\{W_i, i\in[1:n]\}$, each of which is modelled as a vector over $\mathbb F$.
\item For each receiver $j$, a set $D(j)\subseteq {\cal W}$ denoting the set of messages demanded by the receiver $j$.
\item For each receiver $j$, a set $S(j)\subseteq {\cal W}\backslash D(j)$ denoting the set of side-information messages available at the $j^{th}$ receiver.
\end{itemize}
This general class of index coding problems is known as \textit{groupcast} index coding problems. 
\begin{definition}[Index code of symmetric rate $R$]
An index code of symmetric rate $R$ for a given index coding problem consists of an encoding function 
\[
\mathbb{E}:\underbrace{\mathbb F^{LR}\times\mathbb F^{LR}\times...\times\mathbb F^{LR}}_{n~\text{times}}\rightarrow {\mathbb F}^L,
\]
for some $L\geq 1,$ mapping the $n$ $LR$-length message vectors ($W_i\in{\mathbb F}^{LR}$) to some $L$-length codeword which is broadcast through the channel, as well as decoding functions 
\[
\mathbb{D}_j:{\mathbb F}^L\times\underbrace{\mathbb F^{LR}\times\mathbb F^{LR}\times...\times\mathbb F^{LR}}_{|S(j)|~\text{times}}\rightarrow \underbrace{\mathbb F^{LR}\times\mathbb F^{LR}\times...\times\mathbb F^{LR}}_{|D(j)|~\text{times}}
\]
at the receivers $j=[1:T],$ mapping the received codeword and the side-information messages to the demanded messages $D(j)$, i.e., 
\[
{\mathbb D}_j\left(\mathbb{E}(W_1,...,W_n),S(j)\right)=D(j),~\forall~j\in[1:T].
\]
\end{definition}

\begin{remark}
We could in general have different rates for different messages, but in this paper we restrict our attention to symmetric rates. Therefore any rate referred to in this paper is the symmetric rate.
\end{remark}
\begin{definition}[Achievable rates and rate $R$ feasibility]
For a given index coding problem, a rate $R$ is said to be \textit{achievable} if there exists an index code of rate $R$, and the index coding problem is said to be \textit{rate} $R$ \textit{feasible}. 
\end{definition}
\begin{definition}[Scalar index codes and linear index codes]
If a rate $R=1/L$ is achievable, the associated index code is a \textit{scalar index code of length} $L$. If the encoding and decoding functions are linear, then we have a \textit{linear index code}. 
\end{definition}

If we have a linear index code of rate $R$, then we can represent the encoding function as follows.
\[
{\mathbb E}(W_1,W_2,...,W_n)=\sum_{i=1}^nV_iW_i,
\]
where each $V_i$ is a $L\times LR$ matrix with elements from $\mathbb{F}$. In the case of scalar linear index coding, we have $LR=1$. Finding a scalar linear index code of length $L$ (i.e., with a feasible rate $1/L$) is equivalent to finding an assignment of these $L$-length vectors $V_i$s to the $n$ messages such that the receivers can all decode their demanded messages, i.e., 
\[
{\mathbb D}_j\left(\sum_{i=1}^nV_iW_i,S(j)\right)=D(j),~\forall~j\in[1:T].
\]
In the case of a scalar linear index code, the encoding function $\mathbb E$ fixes the vectors assigned to the messages. However not every vector assignment is a valid index coding encoding function, as the decoding functions may not exist. In a number of proofs in this paper, we start with some vector assignment and show that it leads to an encoding function of a valid index code. Therefore, in such proofs we refer to the initial vector assignment as a valid encoding function $\mathbb E$ by abusing the notation.
\begin{remark}
We restrict our attention to scalar linear index codes for the rest of this paper. However we believe that our results can be extended to vector linear index codes as well.
\end{remark}
\subsection{Modelling unavailable side-information as interference}
\begin{definition}[Interfering sets and messages, conflicts]
For some receiver $j$ and for some message $W_k \in D(j)$, let $Interf_k(j)\triangleq {\cal W}\backslash({W_k\cup S(j)})$ denote the set of messages (except $W_k$) not available at the receiver $j$. The sets $Interf_k(j), \forall k$ are called the \textit{interfering sets at receiver} $j$. If receiver $j$ does not demand message $W_k$, then we define $Interf_k(j)\triangleq\phi$. If a message $W_i$ is not available at a receiver $j$ demanding at least one message $W_k\neq W_i$, then $W_i$ is said to \textit{interfere at receiver} $j$, and $W_i$ and $W_k$ are said to be \textit{in conflict}.
\end{definition}

For a set of vertices $A\subseteq {\cal W}$, let $V_{\mathbb E}(A)$ denote the vector space spanned by the vectors assigned to the messages in $A$, under the specified encoding function $\mathbb E$. 
If $A=\phi$, we define $V_{\mathbb E}(A)$ as the zero vector. 
\begin{definition}[Resolved conflicts]
For a given assignment of vectors to the messages (or equivalently, for a given encoding function $\mathbb E$), we say that \textit{conflicts within a subset} ${\cal W}'\subseteq{\cal W}$ \textit{are resolved}, if 
\begin{align}
\nonumber
V_k\notin &V_{\mathbb E}(Interf_k(j)\cap{\cal W}')\\
\label{resolvedconflicts}
 &\forall W_k \in {\cal W}', \forall~\text{receivers}~j\in[1:T],
\end{align}
where $V_k$ is the vector assigned to $W_k$ under the encoding function $\mathbb E$. If (\ref{resolvedconflicts}) holds for ${\cal W}'={\cal W},$ then \textit{all the conflicts} in the given index coding problem \textit{are said to be resolved}.
\end{definition}
We now state a simple lemma, rephrased from \cite{MCJ1}, which is easily proved. 
\begin{lemma}
\label{lemmasuccessdecoding}
For any encoding function $\mathbb E$, successful decoding at the receivers is possible if and only if all the conflicts are resolved.
\end{lemma}
\begin{IEEEproof}
Let $V_k$ be the vector assigned to $W_k$ under the encoding function $\mathbb E$.

\textit{If part:} At all receivers $j$, consider that we have $V_k\notin V_{\mathbb E}(Interf_k(j)),~\forall k\in[1:n]$. What is received at a receiver $j$ is the codeword $\sum_{i=1}^nV_iW_i$, from which it wants to obtain $W_k$. As receiver $j$ can always subtract the contribution from the side-information messages from $\sum_{i=1}^nV_iW_i$, it only remains to be shown that $W_j$ can be decoded from 
\begin{align}
\label{eqn1}
\sum_{i:W_i\notin S(j)}V_iW_i=V_kW_k+\sum_{i:W_i\in Interf_k(j)}V_iW_i,
\end{align}
where the equality is because 
\[
\{W_i\notin S(j)\}=\{W_i\in Interf_k(j)\}\cup\{W_k\}.
\] 
Because of the assumption that $V_k\notin V_{\mathbb E}(Interf_k(j))$, receiver $j$ can get $W_k$ from (\ref{eqn1}). 

\textit{Only if part:} Consider now that there is some receiver $j$ and some message $k$ such that $V_k\in V_{\mathbb E}(Interf_k(j)).$ Clearly, solving for $W_k$ from (\ref{eqn1}) does not lead to an unique solution. Hence decoding fails. This concludes the proof.
\end{IEEEproof}

By Lemma \ref{lemmasuccessdecoding}, it should also be clear that if there is an assignment of $L$-length vectors $V_i$s to the messages $W_i$s such that the condition in Lemma \ref{lemmasuccessdecoding} is satisfied, then these vectors naturally define an index code of length $L$ for the given index coding problem. 

\section{A Relook at Feasibility of rate $\frac{1}{2}$}
\label{sec3}
\subsection{Alignment and conflict graphs of \cite{Jaf}}
In \cite{Jaf}, the authors defined the notions of \textit{alignment graph }and \textit{conflict graph} whose properties were used to characterise index coding problems for which rate $\frac{1}{2}$ is feasible. Both of these graphs have the same vertex set, which is the set of messages $\cal W$. 
\begin{definition}[Alignment graph and alignment sets - \cite{Jaf}]
In the alignment graph, the vertices $W_i$ and $W_j$ are connected by an edge (called an \textit{alignment edge}, shown in our figures by a solid edge) when the messages $W_i$ and $W_j$ are not available at a receiver demanding a message other than $W_i$ and $W_j$. A connected component of the alignment graph is called an \textit{alignment set}.
\end{definition}
It is easy to see that the alignment sets define a partition of the alignment graph. Also, the messages in $Interf_k(j),$ for all messages $k$ at all receivers $j$ are fully connected in the alignment graph. 


\begin{definition}[Conflict Graph - \cite{Jaf}]
In the conflict graph, $W_i$ and $W_j$ are connected by an edge (called an \textit{conflict edge}, shown by a dotted edge) if $W_i$ is not available at a receiver demanding $W_j$, or $W_j$ is not available at a receiver demanding $W_i$. 
\end{definition}

\subsection{Capturing interference in conflict hypergraphs}
\label{conflicthypergraphsubsec}

\begin{figure*}[ht]
  \centering
  \subfigure[]{\label{fig:conflictgraphsame}\includegraphics[height=1in]{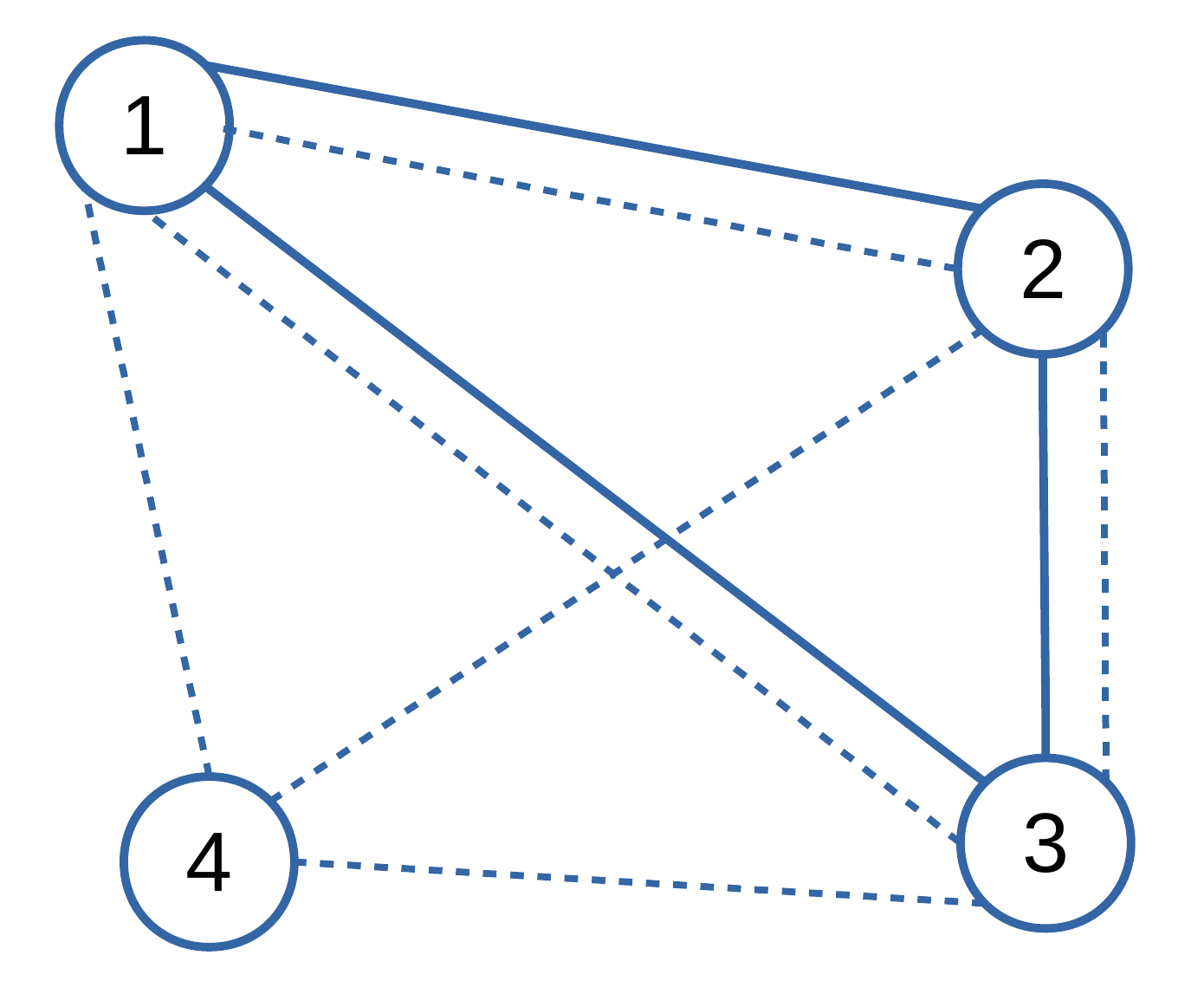}} 
  \hspace{0.2in}
  \subfigure[]{\label{fig:conflicthyp1}\includegraphics[height=1in]{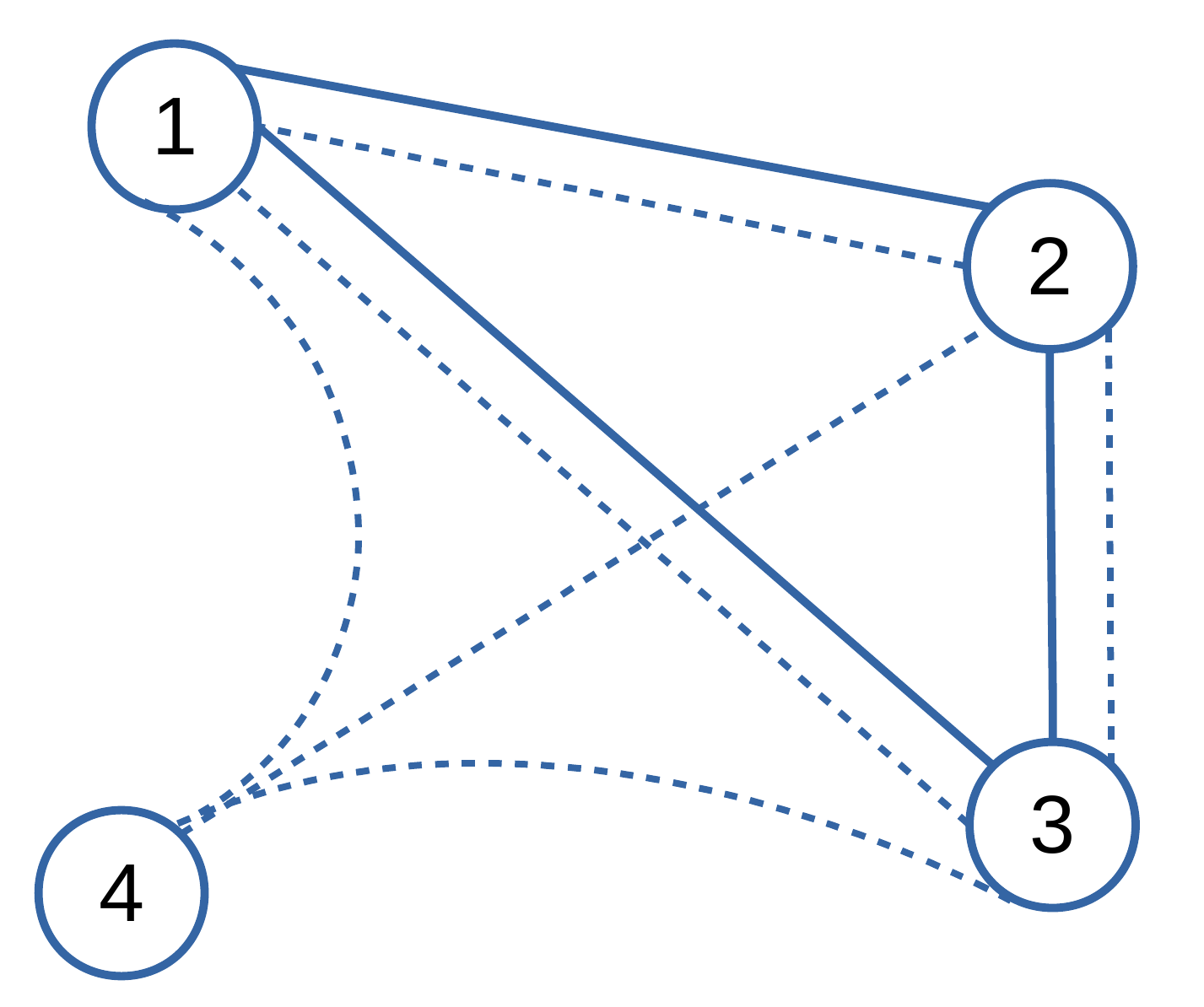}}
    \hspace{0.2in}
  \subfigure[]{\label{fig:conflicthyp2}\includegraphics[height=1in]{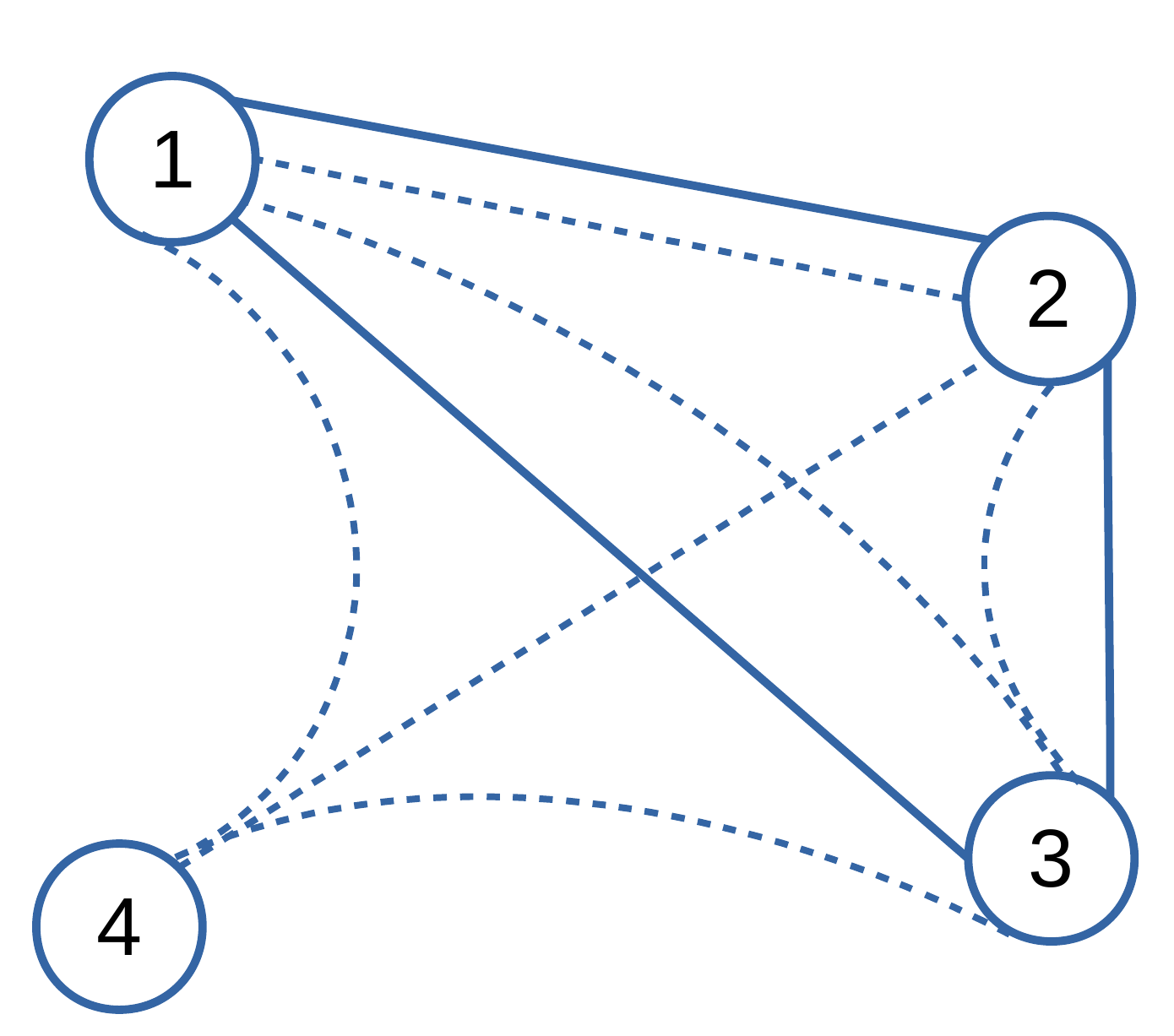}}
  \vspace{-0.1in}
  \caption{(a) Alignment and conflict graphs for the two index coding problems discussed in Example \ref{conflictissueexm}. Edges in alignment graph are shown in solid lines and those in conflict graph are shown in dotted lines. (b) Conflict hypergraph for the first index coding problem in Example \ref{conflictissueexm}. The hyperedges are $\{1,3\}, \{2,1\}, \{3,2\}, \{4,\{1,2,3\}\}$. (b) Conflict hypergraph for the second problem. The hyperedges are $\{2,1\}, \{3,\{1,2\}\}, \{4,\{1,2,3\}\}$.}
 \label{fig:conflictgraph}
\end{figure*}

The following example illustrates that the conflict graph definition does not capture the \textit{directionality} of the conflicts. 
\begin{example}
\label{conflictissueexm}
Consider two single unicast index coding problems with four messages. In the first problem, the interfering sets are as follows $Interf_1(1)=W_3,Interf_2(2)=W_1,Interf_3(3)=W_2, Interf_4(4)=\{W_1,W_2,W_3\}$. In the second problem, $Interf_1(1)=\phi,Interf_2(2)=W_1,Interf_3(3)=\{W_1,W_2\}, Interf_4(4)=\{W_1,W_2,W_3\}.$ All other interfering sets corresponding to the receivers are empty. These two problems have the same alignment and conflict graphs (see Fig.\ref{fig:conflictgraphsame}). In particular, the conflict graphs are the same because the definition of the conflict graph does not model the directionality of the absent side-information messages. Note that they both have different solutions. We leave it to the reader to check that the first problem is rate $\frac{1}{2}$ feasible, while the second problem is not rate $\frac{1}{2}$ feasible, but has a rate $\frac{1}{3}$ solution. 
\end{example}

To overcome this issue, we define the \textit{conflict hypergraph} as follows.
\begin{definition}[Conflict hypergraph]
The conflict hypergraph is an undirected hypergraph with vertex set $\cal W$ (the set of messages), and its hyperedge set defined as follows. 
\begin{itemize}
\item For any receiver $j$ demanding any message $W_k$, $W_k$ and $Interf_k(j)$ are connected by a hyperedge, which is denoted by $\{W_k,Interf_k(j)\}$.
\end{itemize}
%
\end{definition}

For example, the two problems presented in Example \ref{conflictissueexm} are now represented using different conflict hypergraphs in Fig. \ref{fig:conflicthyp1} and Fig. \ref{fig:conflicthyp2} (the alignment graphs remains the same). Note that even though this definition for the conflict hypergraph does not explicitly contain direction, the directionality of interference seen by any receiver is modelled correctly whenever the number of interfering messages is more than one. This is sufficiently general as we see in the following Lemma. 

\begin{lemma}
Suppose two index coding problems, denoted by ${\mathbb I}_1$ and ${\mathbb I}_2,$ are modelled by the same conflict hypergraph. Then any index coding solution for ${\mathbb I}_1$ is an index coding solution for ${\mathbb I}_2$.
\end{lemma}
\begin{IEEEproof}
Let $\mathbb{E}$ be the encoding function of the given index code for ${\mathbb I}_1$. Let $V_k$ be the vector assigned to $W_k$. By Lemma \ref{lemmasuccessdecoding}, we must have that for $V_k\notin V_{\mathbb E}(Interf_k(j)),~\forall k\in[1:n],~\forall j\in[1:T]$ in ${\mathbb I}_1$. Now assume that the same index code is used for ${\mathbb I}_2$. For any receiver $j$ and message $W_{k}$ with $|Interf_{k}(j)|\geq 2$ in ${\mathbb I}_2$, the conflict hyperedge $\{W_{k},Interf_{k}(j)\}$ present in the conflict hypergraph of ${\mathbb I}_2$ is present also in that of ${\mathbb I}_1$ as both ${\mathbb I}_1$ and ${\mathbb I}_2$ have the same conflict hypergraph. In other words, message $Interf_{k}(j)$ is an interfering set of a receiver $j$ in ${\mathbb I}_1$ also and hence we must have $V_k\notin V_{\mathbb E}(Interf_{k}(j))$, which means that $W_{k}$ is recoverable at $j$ in ${\mathbb I}_2$ also. 

The only case left to check is when $|Interf_{k}(j)|=1$ in ${\mathbb I}_2$ for some receiver $j$ and message $W_{k}$. Let us assume that $Interf_{k}(j)=W_{i}.$ Because ${\mathbb I}_1$ and ${\mathbb I}_2$ share the same conflict hypergraphs, this conflict (in some direction) is present in ${\mathbb I}_2$ also. By assignment $\mathbb E$, we must have that $V_i$ and $V_k$ are linearly independent, which ensures that this conflict is resolved in ${\mathbb I}_2$ also, irrespective of its directionality.
\end{IEEEproof}

\subsection{A new lemma and its application: Rate half feasibility condition from \cite{Jaf}}
Towards showing a necessary and sufficient condition for rate $\frac{1}{2}$ feasibility, the following definition for \textit{internal conflicts} was given in \cite{Jaf}.
\begin{definition}[Internal conflict \cite{Jaf}]
A conflict between two messages within an alignment set is called an \textit{internal conflict}.
\end{definition}

The following theorem was proved in \cite{Jaf} on rate $\frac{1}{2}$ feasible index coding problems. 
\begin{theorem}
\label{ratehalfthm}
An index coding problem is rate $\frac{1}{2}$ feasible if and only if there are no internal conflicts.
\end{theorem}

The following lemma plays a crucial role in our proof of Theorem \ref{ratehalfthm}, which is included for the sake of completeness as it is not available in its complete form in prior literature.  We will also see in Subsection \ref{rateonethirdfeasibleproblems} that it is central to proving our main theorem which deals with the feasibility of rate $\frac{1}{3}$.
\begin{lemma}
\label{intersectingKdimspaces}
Let $U_1,U_2,...,U_N$ be $N$ sets of vectors, such that $dim(sp(U_j\cap U_{j+1}))=K, \forall j\in[1:N-1]$. Then the space spanned by $\cup_{j=1}^NU_j$ has dimension $K$ if and only if each $U_j$ spans a vector space of dimension $K$.
\end{lemma}
\begin{IEEEproof}
\textit{If part:} Suppose each $U_j$ spans a vector space of dimension $K$, and $dim(sp(U_j\cap U_{j+1}))=K, \forall j\in[1:N-1].$ Clearly, all the vector spaces $sp(U_j)$s are exactly equal, and thus $dim(sp(\cup_{j=1}^NU_j))=K.$  

\textit{Only if part:} Suppose $dim(sp(\cup_{j=1}^NU_j))= K$ and 
\begin{equation}
\label{eqn3}
dim(sp(U_j\cap U_{j+1}))=K, \forall j\in[1:N-1].
\end{equation}
By (\ref{eqn3}), $dim(sp(U_j))\geq K, \forall j$. If some $dim(sp(U_j))>K$, then $dim(sp(\cup_{j=1}^NU_j))>K$, and we have a contradiction. Thus, we must have that  $dim(sp(U_j))=K, \forall j$.
\end{IEEEproof}

We now use Lemma \ref{intersectingKdimspaces} and Lemma \ref{lemmasuccessdecoding} to prove Theorem \ref{ratehalfthm}.
\begin{IEEEproof}[Proof of Theorem \ref{ratehalfthm}]
Corresponding to any vertex $W_k$ in the alignment graph, let $Align(k)$ denote the alignment set it belongs to (this is unique as the alignment sets partition the alignment graph). We first note that in any (scalar linear) index coding scheme for the given problem, all the vertices must be assigned some non-zero vectors (zero vector cannot be assigned to any message as this means that the message cannot be decoded by any receiver).

\textit{If part:} Suppose that there are no internal conflicts. We assume a large field $\mathbb{F}$. For each alignment set, we independently generate a random $2\times 1$ vector over $\mathbb F$ and assign it to the vertices of the alignment set. Because of random generation, we can assume that any assigned vector is non-zero and any two assigned vectors are linearly independent with high probability (whp). Let $\mathbb E$ denote the associated encoding function and $V_k$ denote the vector assigned to vertex $W_k$. Since there are no internal conflicts, we only have to check conflicts between alignment sets. For any vertex $W_k$, the set $Interf_k(j)$ for any receiver $j$ which demands $W_k$ belongs to a $(i)$ different and $(ii)$ unique alignment set than $Align(k)$ ($(i)$ is because there are no internal conflicts, $(ii)$ is because all the messages in $Interf_k(j)$ must be in the same alignment set). Since any two alignment sets get independent vectors (whp), we have that $V_k\notin V_{\mathbb E}(Interf_k(j))$, and the same argument is true for all receivers $j$ and all messages $W_k$. Hence this assignment of vectors ensures successful decoding by Lemma \ref{lemmasuccessdecoding}.

\textit{Only if part:} Suppose that there is some internal conflict (represented in the conflict graph as an edge between node $k'$ and node $k$) in an alignment set $Align(k)$. Because $k',k$ are part of the same alignment set (connected component) $Align(k)$, there lies a path from $k'$ to $k$, given by an ordered set $\{k',i_1,i_2,...,i_{N-1},k\}$, such that every adjacent pair of elements ($\{k',i_1\}$, etc.) belong to the interfering set of some receiver. 

In some assignment corresponding to a rate $\frac{1}{2}$ solution, let $V_{k'},V_{i_1},...,V_{i_{N-1}},V_k$ be the non-zero vectors assigned to the vertices $\{k',i_1,i_2,...,i_{N-1},k\}$. We define the sets, $U_1\triangleq\{V_{k'},V_{i_1}\},~U_l\triangleq\{V_{i_{l-1}},V_{i_{l}}\},\forall l\in[2:N-1]$ and $U_N\triangleq\{V_{i_{N-1}},V_{k}\}$. 

\begin{figure}[ht]
\centering
\includegraphics[width=2.2in]{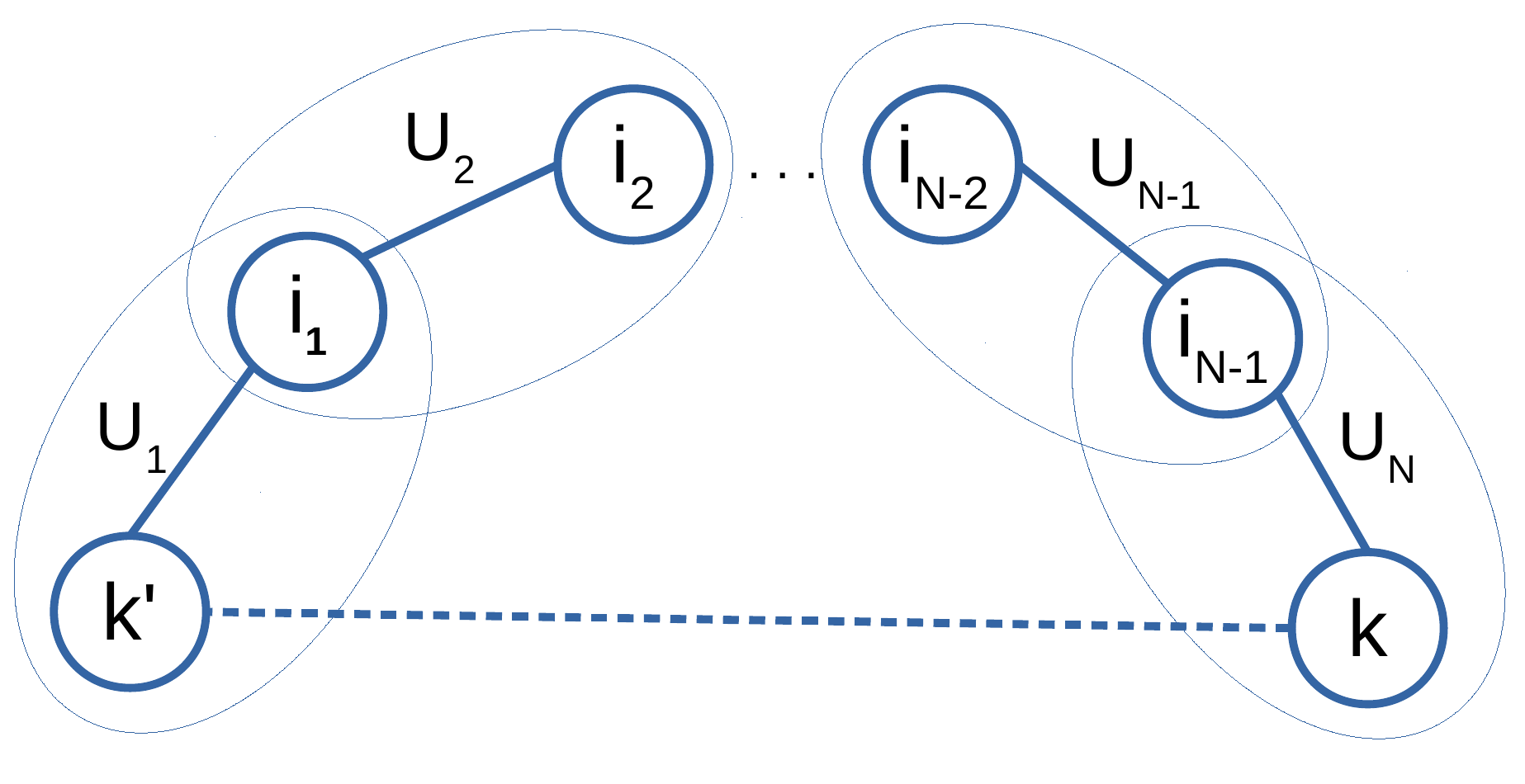}
\caption{Since $k'$ and $k$ are in conflict, they must be assigned linearly independent vectors. This requires that at least one of the sets $U_1, \ldots, U_N$ has to be two dimensional.}	
\label{fig:ratehalfconflict}	
\end{figure}

Suppose some $dim(sp(U_l))=2$ for some $l$. Then a receiver $j$ (at which the messages corresponding to $U_l$ are unavailable) `sees' an interfering space of dimension $2$. Thus we need to assign a vector linearly independent from $U_l$ (which itself has dimension $2$) to the corresponding demanded message of receiver $j$. This can be possible only if the assigned vectors are of length at least $3$, i.e. the rate can be at most $\frac{1}{3}$. 

Therefore, for a rate $\frac{1}{2}$ index coding assignment, all $U_l$ should spanning a space of dimension $1$. Thus we have $dim(sp(U_l))=1,\forall l$ and $dim(sp(U_l\cap U_{l+1}))=1,\forall l\in[1:N-1].$ By Lemma \ref{intersectingKdimspaces}, we should thus have $dim(sp(\cup_{l=1}^NU_l))=1.$ 

However $k'$ and $k$ are in conflict, which means that they should be assigned linearly independent vectors, i.e., $dim(sp(\{V_{k'},V_k\}))=2$ which means that $dim(sp(\cup_{l=1}^NU_l))>1$. Thus there is a contradiction and thus any internal conflicts forces the rate to be less than $\frac{1}{2}$. This concludes the proof.
\end{IEEEproof}
\section{Feasibility of rate $\frac{1}{3}$}
From Section \ref{sec3}, the following is clear.
\begin{itemize}
\item If there are no conflicts (not even conflicts between two alignment sets) in the alignment graph, rate $1$ is achievable (this is the case when any receiver demanding a message has all the other messages as side-information).
\item For rate $1$ infeasible index coding problems, rate $\frac{1}{2}$ is feasible if and only if there are no internal conflicts. 
\end{itemize}
Towards obtaining our main result, which characterises a class of index coding problems which are rate $\frac{1}{3}$ feasible, we first give a prior known necessary condition for feasibility of rate $\frac{1}{3}$.

\subsection{A known necessary condition for rate $\frac{1}{3}$ feasibility}
Suppose there are $4$ messages $\{W_{i_k}, k=1,..,4\}$ such that message $W_{i_k}$ is demanded by some receiver (say receiver $j_k$) and $\{W_{i_{k'}}:k'<k\}\subseteq Interf_{i_k}(j_k).$ Following \cite{Jaf}, we call such a set of messages $\{W_{i_k}, k=1,..,4\}$  as an \textit{acyclic subset of messages of size} $4$. The following theorem can be obtained from the results in \cite{BBJK}. 
\begin{theorem}
\label{thmMAIS}
An index coding problem which is rate $\frac{1}{3}$ feasible cannot have an acyclic subset of messages of size $4$.
\end{theorem}
%

\begin{figure}[ht]
\centering
\includegraphics[width=2in]{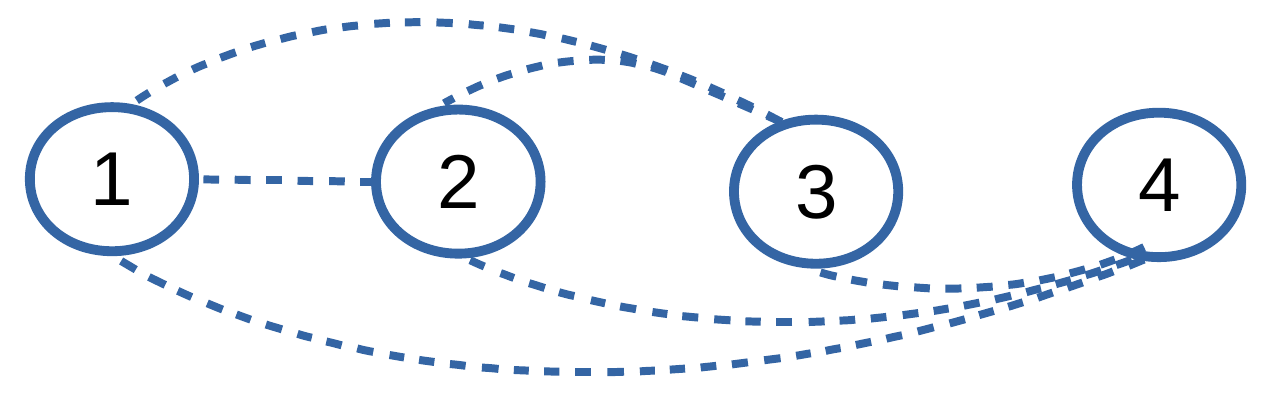}
\caption{From the structure of the conflict hypergraph, it can be seen that vectors $\{V_1, \ldots, V_4\}$ assigned to the four messages must be linearly independent.}	
\label{fig:maisconflict}	
\end{figure}

\begin{figure}[h]
  \centering
  \subfigure[]{\label{fig:infeasible} \includegraphics[width=1.7in]{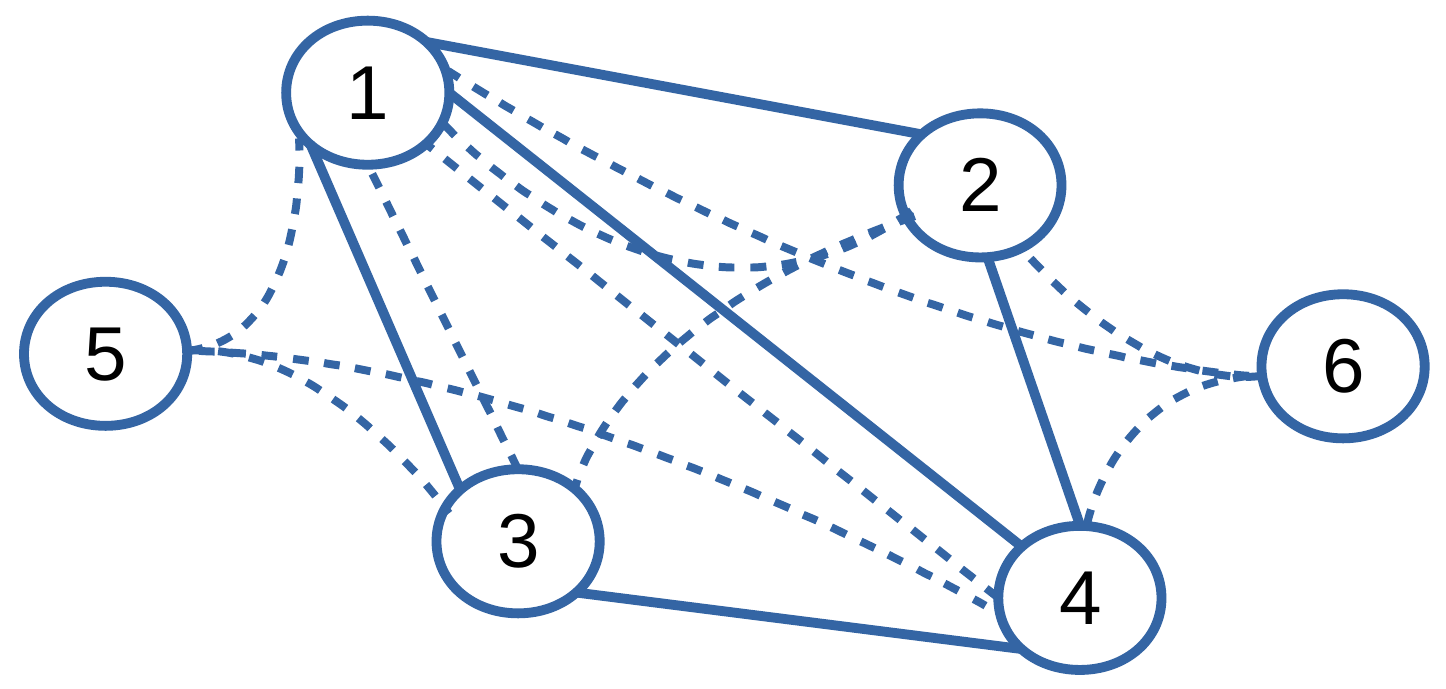}} 
  \hspace{0.1in}
  \subfigure[]{\label{fig:feasible} \includegraphics[width=1.4in]{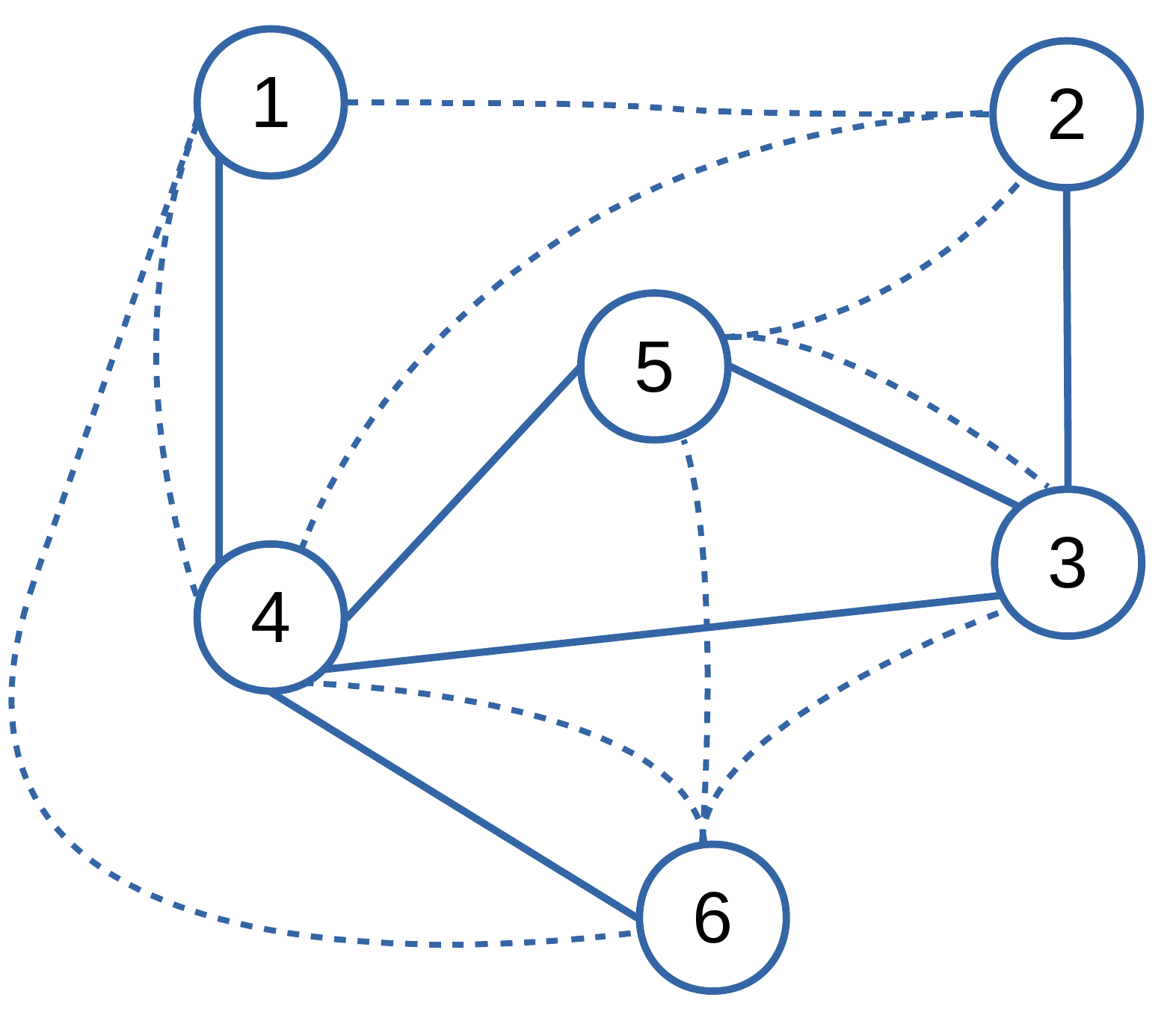}}
  \vspace{-0.1in}
  \caption{(a) Alignment graph and conflict hypergraph for the problem in Example \ref{ex:onethirdinfeasible}, which is rate $\frac{1}{3}$ infeasible. (b)  Alignment graph and conflict hypergraph for the problem in Example \ref{ex:onethirdfeasible}, which is rate $\frac{1}{3}$ feasible.}
  \label{fig:rateonethird}
\end{figure}

The following example however shows that Theorem \ref{thmMAIS} is not a sufficient condition for rate $\frac{1}{3}$ feasibility.

\begin{example}
\label{ex:onethirdinfeasible}

Consider a single unicast index coding problem with six messages. The interfering sets are as follows: $Interf_1(1)=W_4,Interf_2(2)= \{W_1, W_3\}, Interf_3(3)=W_1, Interf_4(4)=\phi, Interf_5(5)=\{W_1,W_3,W_4\}, Interf_6(6) = \{W_1,W_2,W_4\}$. The alignment graph and the conflict hypergraph corresponding to the problem are given in Fig. \ref{fig:infeasible}. It can be seen that the conflict hypergraph of this problem does not have an acyclic subset of messages of size $4$. If we assume that the problem is rate $\frac{1}{3}$ feasible, then it has to be necessarily true that both $sp(\{V_1, V_3, V_4\})$ and $sp(\{V_1, V_2, V_4\})$ are two dimensional. We leave it to the reader to check that $sp(\{V_1, V_2, V_3\})$ must also be two dimensional. However, the conflict between the messages $W_1, W_2$ and $W_3$ are resolved only if $sp(\{V_1, V_2, V_3\})$ is three dimensional. Thus, the problem is rate $\frac{1}{3}$ infeasible.
\end{example}

\subsection{A known sufficient condition for rate $\frac{1}{3}$ feasibility}
In \cite{Jaf}, the following theorem about the achievability of rate $\frac{1}{3}$ was proved (this is a special case of Corollary 9 of \cite{Jaf}). 
\begin{theorem}
\label{thmnocyclesforks}
Consider a rate $\frac{1}{2}$ infeasible index coding problem with no acyclic subset of size $4.$ If none of its alignment sets have both forks (a \textit{fork} is a vertex connected by three or more edges) and cycles,  then the index coding problem is rate $\frac{1}{3}$ feasible.
\end{theorem}

As mentioned in \cite{Jaf}, the condition that there are no alignment sets with both forks and cycles of length $3$ means that there is no interfering set $Interf_k(j)$ at any receiver of size $\geq 4$, as such a set would mean that there is both a cycle and fork within an alignment set (since the messages in $Interf_k(j)$ are fully connected in the alignment set). Therefore Theorem \ref{thmnocyclesforks} characterises a rather limited class of index coding problems which are $\frac{1}{3}$ feasible. Example \ref{ex:onethirdfeasible} shows an index coding problem which does not satisfy the conditions of Theorem \ref{thmnocyclesforks} but is rate $\frac{1}{3}$ feasible. 

\begin{example}[Example to illustrate that the condition in Theorem \ref{thmnocyclesforks} is not necessary] \label{ex:onethirdfeasible}
Consider a single unicast index coding problem with six messages. The interfering sets of the problem are as follows: $Interf_1(1)=\{W_4, W_6\}, Interf_2(2)= \{W_1, W_4\}, Interf_3(3)=\phi, Interf_4(4)=\phi, Interf_5(5)=\{W_2,W_3\}, Interf_6(6) = \{W_3,W_4,W_5\}$. The alignment graph and the conflict hypergraph corresponding to the problem are given in Fig. \ref{fig:feasible}. It can be seen that the alignment graph of this problem has both forks and a cycle. Consider $3$ linearly independent $3 \times 1$ vectors $V_1, V_2, V_3$. We note that the following assignment of vectors resolves all conflicts: 
(i) vector $V_1$ to messages $W_1$ and $W_3$, (ii) vector $V_2$ to messages $W_4$ and $W_5$, (iii) vector $V_3$ to messages $W_2$ and $W_6$. Thus, the problem is rate $\frac{1}{3}$ feasible.
\end{example}

\subsection{Triangular interfering sets and Type-2 alignment sets}
In this subsection, we develop a new framework for studying the rate $\frac{1}{3}$ feasibility of groupcast index coding problems. Towards this, we define the notions of a triangular interfering set and a type-2 alignment set.
\begin{definition}[Triangular Interfering Sets]
A subset ${\cal W}''$ of size three of the set of messages ${\cal W}$ is said to be a \textit{triangular interfering set} if all the messages in ${\cal W}''$ interfere simultaneously at some receiver, and at least two of the messages in ${\cal W}''$ are in conflict.
\end{definition}
\begin{definition}[Adjacent Triangular Interfering Sets]
Two distinct triangular interfering sets ${\cal W}_1$ and ${\cal W}_2$ are said to be \textit{adjacent} if they `meet' at a conflicting edge, i.e.,  ${\cal W}_1\cap{\cal W}_2=\{W_i,W_j\}$ such that $W_i$ and $W_j$ are in conflict. 
\end{definition}
\begin{definition}[Connected triangular interfering sets, Type-2 alignment sets]
\label{type2align}
Two triangular interfering sets ${\cal W}_1$ and ${\cal W}_2$ are said to be \textit{connected} if there exists a \textit{path} (i.e., a sequence) of adjacent triangular interfering sets starting from ${\cal W}_1$ and ending at ${\cal W}_2$. A \textit{type-2 alignment set} is a maximal set of triangular interfering sets which are connected to each other.
\end{definition}
%

\begin{figure}[h]
  \centering
  \subfigure[]{\includegraphics[height=1.2in]{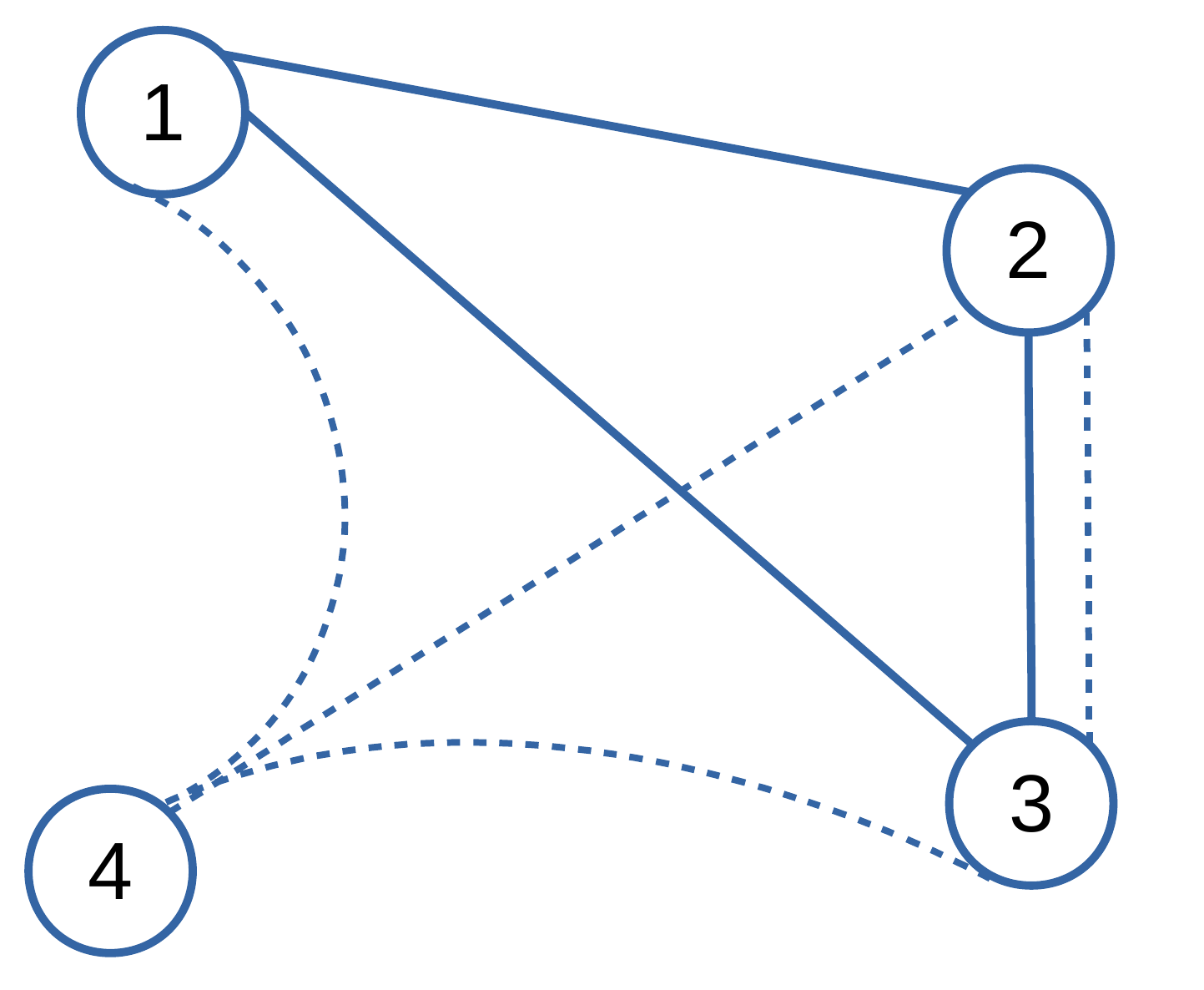}} 
  \hspace{0.2in}
  \subfigure[]{\includegraphics[height=1in]{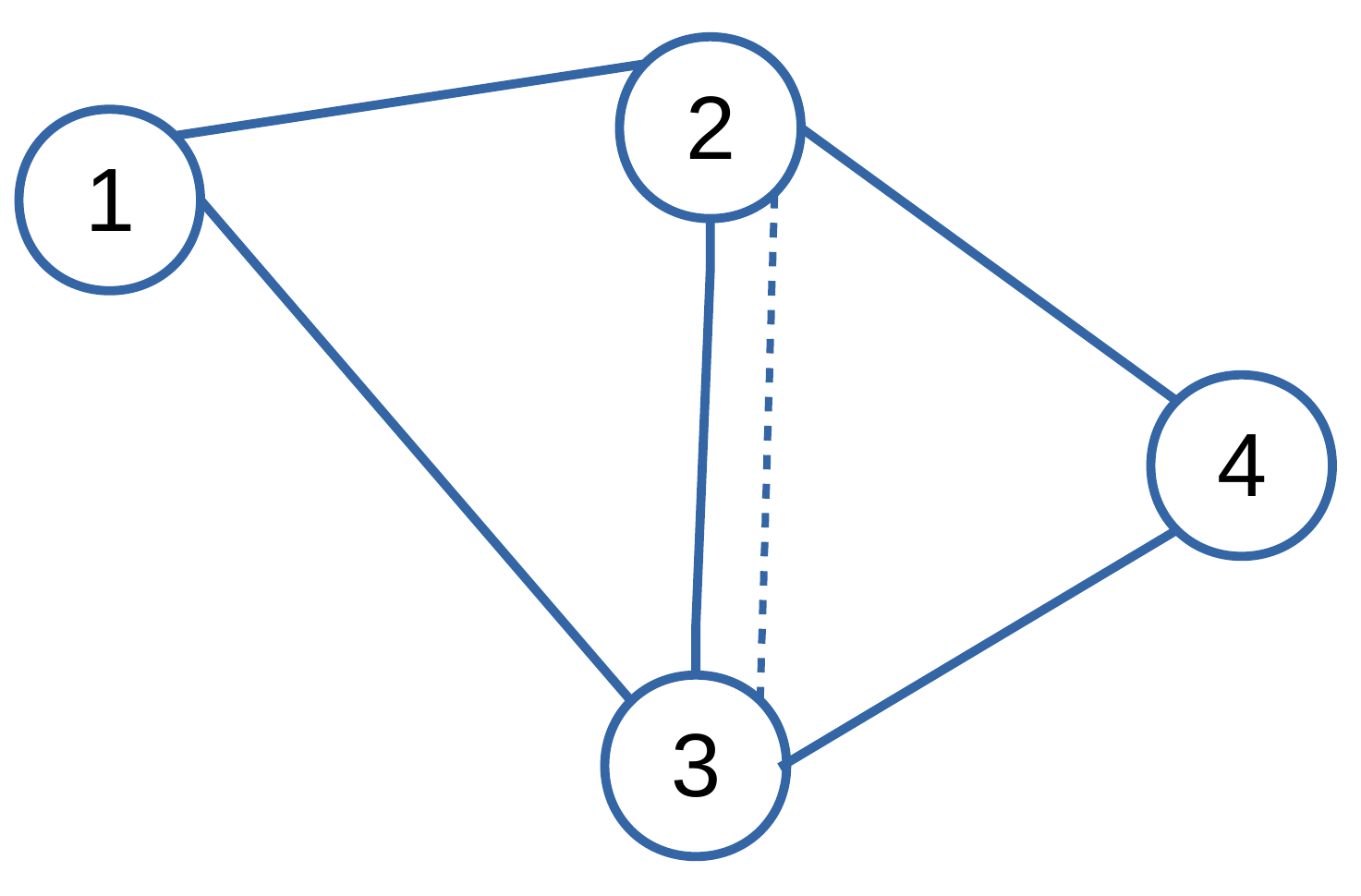}}
    \subfigure[]{\label{fig:type2ex} \includegraphics[height=1.9in]{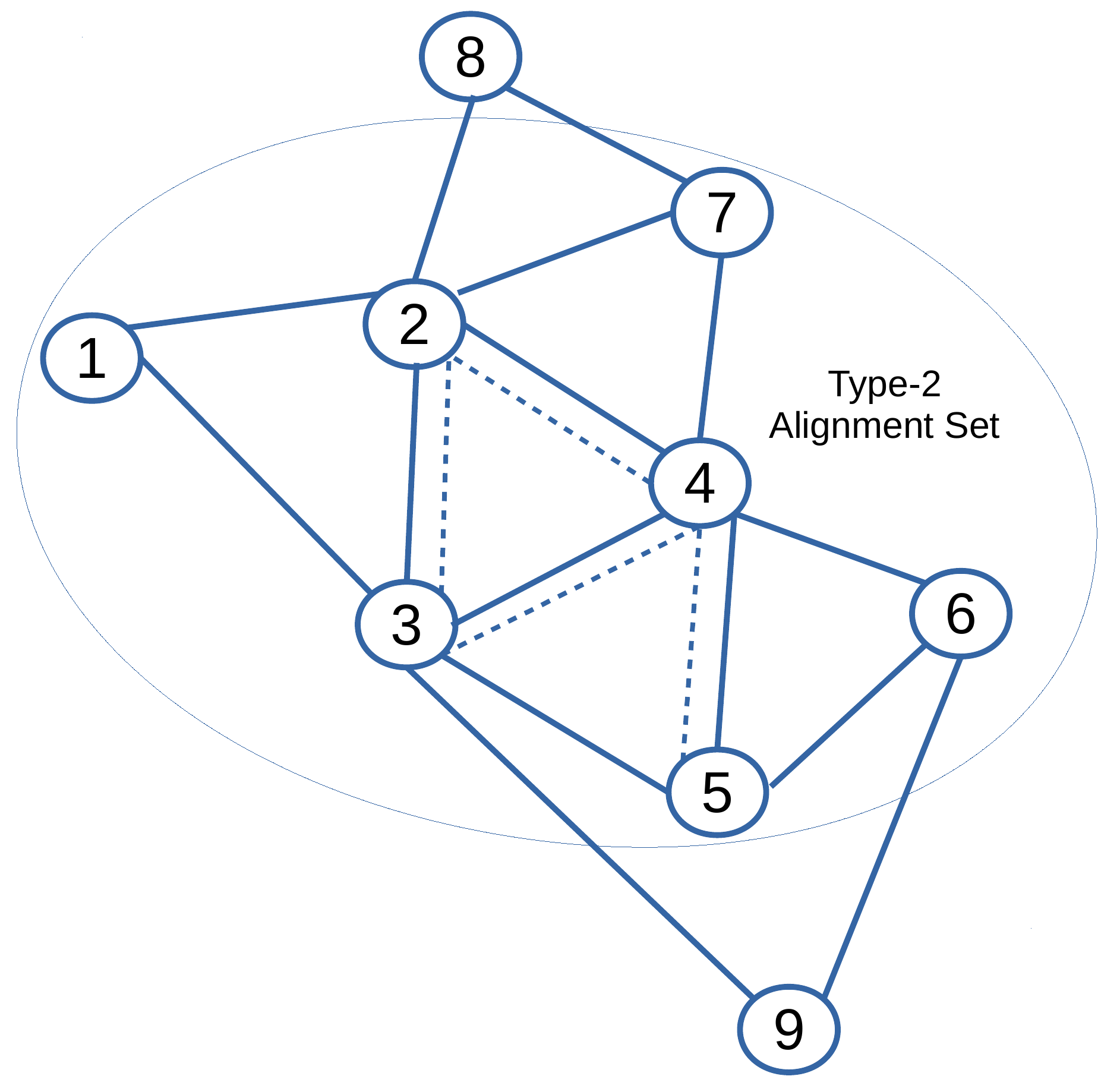}}
  \vspace{-0.1in}
  \caption{(a) Messages $1,2,3$ interfere at a receiver which demands $4$ and messages $2,3$ are in conflict. Thus, $\{1,2,3\}$ form a triangular interfering set. (b) Triangular interfering sets $\{1,2,3\}, \{2,3,4\}$ are adjacent as they both have common conflict edge $\{2,3\}$. The receivers at which these triangular interfering sets interfere are not explicitly shown in the figure. (c) Type-2 alignment set is the subset of messages indicated in the ellipse, the triangular interfering sets include $\{1,2,3\}, \{2,3,4\}, \{3,4,5\}, \{4,5,6\}, \{2,4,7\}$. These $5$ triangular interfering sets are connected.}
  \label{fig:triangleinterfset}
\end{figure}

Examples to illustrate the above definitions are shown in Fig. \ref{fig:triangleinterfset}. Note that the maximality in Definition \ref{type2align} means that we cannot add another triangular interfering set to a type-2 alignment set and still maintain connectivity (as in Definition \ref{type2align}). 

By definition, every type-2 alignment set must be a subset of a (regular) alignment set, and there could be many type-2 alignment sets within any alignment set. Given an index coding problem, we can identify type-2 alignment sets as follows. Within any alignment set, we identify a triangular interfering set of messages (if there is no such set then there is no type-2 alignment set inside that alignment set). Then we repeatedly add triangular interfering sets which are adjacent to the existing connected triangular interfering sets. When we can no longer add such adjacent triangular interfering sets, then we have our type-2 alignment set. 

\subsection{A new stricter necessary condition for rate $\frac{1}{3}$ feasibility}
\label{necessaryconditionrateonethirds}
We now prove a necessary condition for rate $\frac{1}{3}$ feasibility based on the vectors assigned to type-2 alignment sets. This theorem is another application of the key lemma, Lemma \ref{intersectingKdimspaces}. 
\begin{theorem}
\label{type2sets2dim}
In any rate $\frac{1}{3}$ solution to a given index coding problem, all the messages in any type-2 alignment set must be assigned vectors from a vector space of dimension two.
\end{theorem}
\begin{IEEEproof}
Let $\mathbb E$ be the encoding function of a rate $\frac{1}{3}$ solution. Consider a type-2 alignment set ${\cal W}'$ with triangular interfering sets ${\cal W}_i, i\in[1:S].$ 

Suppose for any triangular interfering set ${\cal W}_i$ of ${\cal W}'$, we have  $dim(V_{\mathbb E}({\cal W}_i))=3$. Since all the vertices in ${\cal W}_i$ interfere at some receiver (say, a receiver which requests some other message $W_j$), the message $W_j$ must be assigned a vector which is linearly independent from those assigned to the messages in ${\cal W}_i$, and thus we need at least $4$ linearly independent vectors, and hence the rate has to be $\leq 1/4$. Thus no triangular interfering set ${\cal W}_i$ of ${\cal W}'$ has $dim(V_{\mathbb E}({\cal W}_i))=3$. However, any triangular interfering set ${\cal W}_i$ of ${\cal W}'$ \textit{must} have $dim(V_{\mathbb E}({\cal W}_i))=2$, as ${\cal W}_i$ has a conflict.

Suppose $dim(V_{\mathbb E}({\cal W}'))>2$. Consider three messages $W_{j_1}, W_{j_2},$ and $W_{j_3}$ in ${\cal W}',$ that have been assigned three linearly independent vectors, belonging to some three triangular interfering sets (not necessarily different), ${\cal W}_{i_1}, {\cal W}_{i_2},$ and ${\cal W}_{i_3}$ respectively. We have already shown that we cannot have ${\cal W}_{i_1}={\cal W}_{i_2}={\cal W}_{i_3}$. So at least two of the three triangular interfering sets are different. 

Suppose all three sets ${\cal W}_{i_1}, {\cal W}_{i_2},$ and  ${\cal W}_{i_3}$ are different. Because the three triangular interfering sets are within the same type-2 alignment set, it must be the case that there exists a path consisting of adjacent triangular interfering sets starting from ${\cal W}_{i_1},$ through ${\cal W}_{i_2}$ and upto ${\cal W}_{i_3}.$  Let $N$ be the number of triangular interfering sets on this path (counted as we go along the path; repetitions are counted separately). For $i\in[1:N]$, let $U_i$ denote the set of $3$ vectors assigned to the $i^{th}$ triangular interfering set in this path. Fig. \ref{fig:type2assignment} illustrates this scenario for the type-2 alignment set example shown in Fig. \ref{fig:type2ex}.

\begin{figure}[ht]
\centering
\includegraphics[width=2.6in]{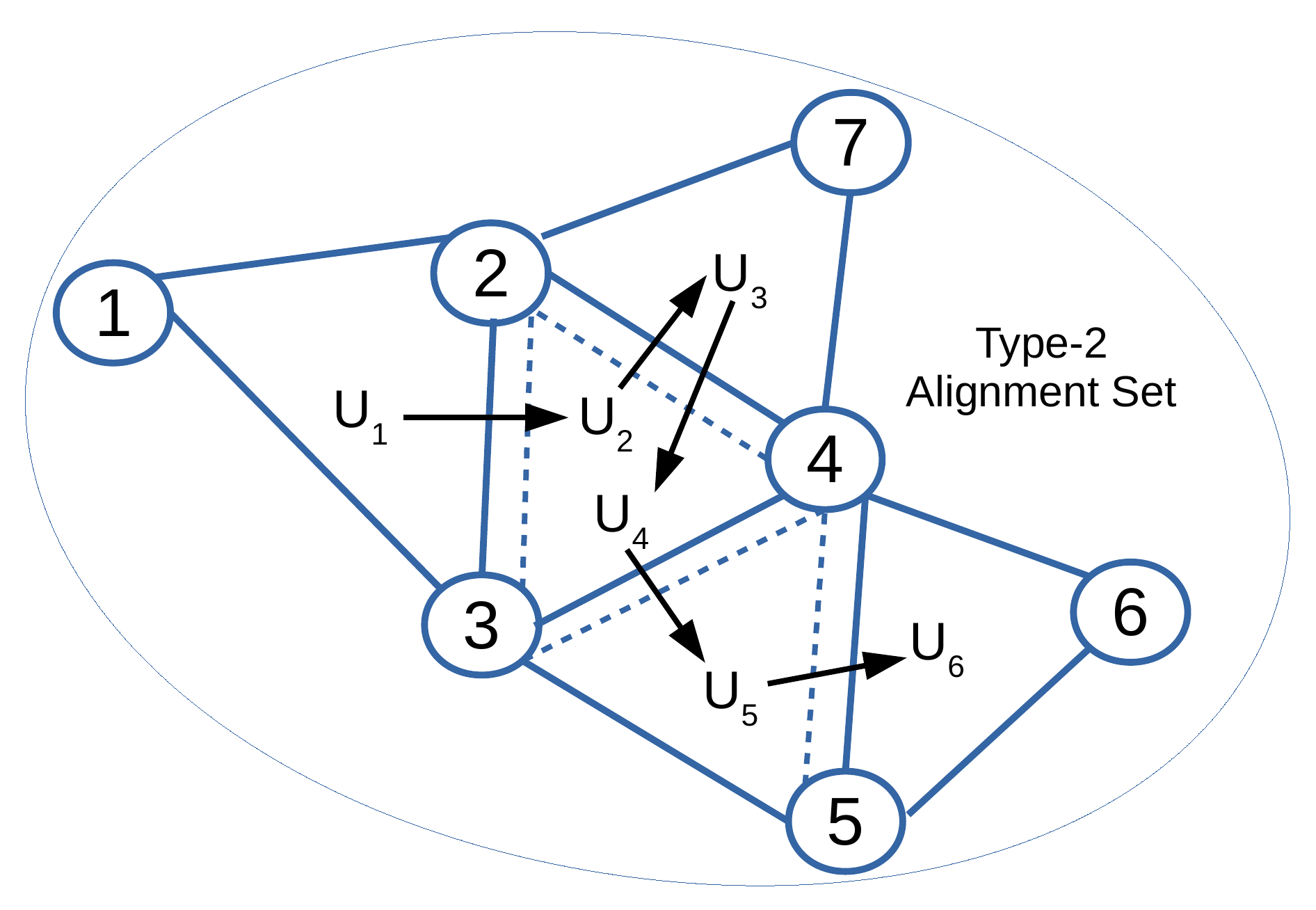}
\caption{Consider messages $1,7,6$. To determine the dimension of the span of these messages, consider the following path of triangular interfering sets: $\{1,2,3\}, \{2,3,4\}, \{2,4,7\}, \{2,3,4\}, \{3,4,5\}, \{4,5,6\}$ and $U_i, 1\leq i \leq 6$ are the sets of $3$ vectors assigned to $i^{th}$ triangular interfering set.}	
\label{fig:type2assignment}	
\end{figure}

By the previous arguments, we have that $dim(sp(U_i))=2, \forall i.$ Also, $dim(sp(U_i\cap U_{i+1}))=2, i\in[1:N-1],$ as the $i^{th}$ and the $(i+1)^{th}$ triangular interfering sets are adjacent by construction of the path. Therefore, by Lemma \ref{intersectingKdimspaces}, it must be the case that $dim(sp(\cup_{i=1}^NU_i))=2$. However, the vector assigned to  the message $W_{j_k}$ belongs to $\cup_{i=1}^NU_i$ for $k=1,2,3$, and according to our assumption the vectors assigned to these three messages are linearly independent vectors. Thus there is a contradiction, which means that we cannot have three messages in three different triangular interfering sets which have been assigned linearly independent vectors. A similar claim can be proved if the three messages come from two different triangular interfering sets. 

Thus, no three messages in a type-2 alignment set can be assigned linearly independent vectors. In other words, any type-2 alignment set ${\cal W}'$ in a rate $\frac{1}{3}$ solution must have $dim(V_{\mathbb E}({\cal W}'))=2$.
\end{IEEEproof}
Theorem \ref{type2sets2dim} is stricter than Theorem \ref{thmMAIS}, as Theorem \ref{thmMAIS} applies only to an acyclic subset of messages of size $4,$ which basically is equivalent to a triangular interfering set. Theorem \ref{type2sets2dim} on the other hand considers a `connected component' of such triangular interfering sets, and is therefore more strict. We leave it to the reader to verify that the problem in Example \ref{ex:onethirdinfeasible}, while `passing' the condition of Theorem \ref{thmMAIS}, `fails' the condition of Theorem \ref{type2sets2dim}.
\subsection{Restricted index coding problems and rate $\frac{1}{2}$ feasibility}
Theorem \ref{type2sets2dim} prescribes that type-2 alignment sets must be `two-dimensional' in a rate $\frac{1}{3}$ code. In this subsection, we give a necessary and sufficient condition for achieving this two-dimensionality. For this purpose, we require the notion of a restricted index coding problem.
\begin{definition}[Restricted Index Coding problem]
Let $\mathbb I$ denote an index coding problem with message set $\cal W$. For some ${\cal W}'\subseteq {\cal W}$, a ${\cal W}'$\textit{-restricted index coding problem} is defined as the index coding problem ${\mathbb I}_{{\cal W}'}$ consisting of 
\begin{itemize}
\item The messages ${\cal W}'$.
\item The subset ${\cal T}_{{\cal W}'}$ (of size ${T}_{{\cal W}'}$) of the receivers of $\mathbb I$ which demand messages in ${\cal W}'$.
\item For each $j\in {\cal T}_{{\cal W}'}$ the demand sets $D_{{\cal W}'}(j)$ and the side-information sets $S_{{\cal W}'}(j)$ are restricted within ${\cal W}'$, i.e., 
\begin{align*}
D_{{\cal W}'}(j)= D(j)\cap{\cal W}'.\\
S_{{\cal W}'}(j)= S(j)\cap{\cal W}'.
\end{align*}
\end{itemize}
\end{definition}

\begin{definition}[`Restricted' versions of alignment graphs, alignment sets, and internal conflicts]
The alignment graph and the alignment sets of the restricted index coding problem ${\mathbb I}_{{\cal W}'}$ are called the ${\cal W}'$-\textit{restricted alignment graph} and ${\cal W}'$-\textit{restricted alignment sets} respectively. A ${\cal W}'$-\textit{restricted internal conflict} is a conflict between any two messages within a restricted alignment set of ${\cal W}'$.
\end{definition}

%
%

\begin{figure}[h]
  \centering
  \subfigure[]{\includegraphics[width=1.7in]{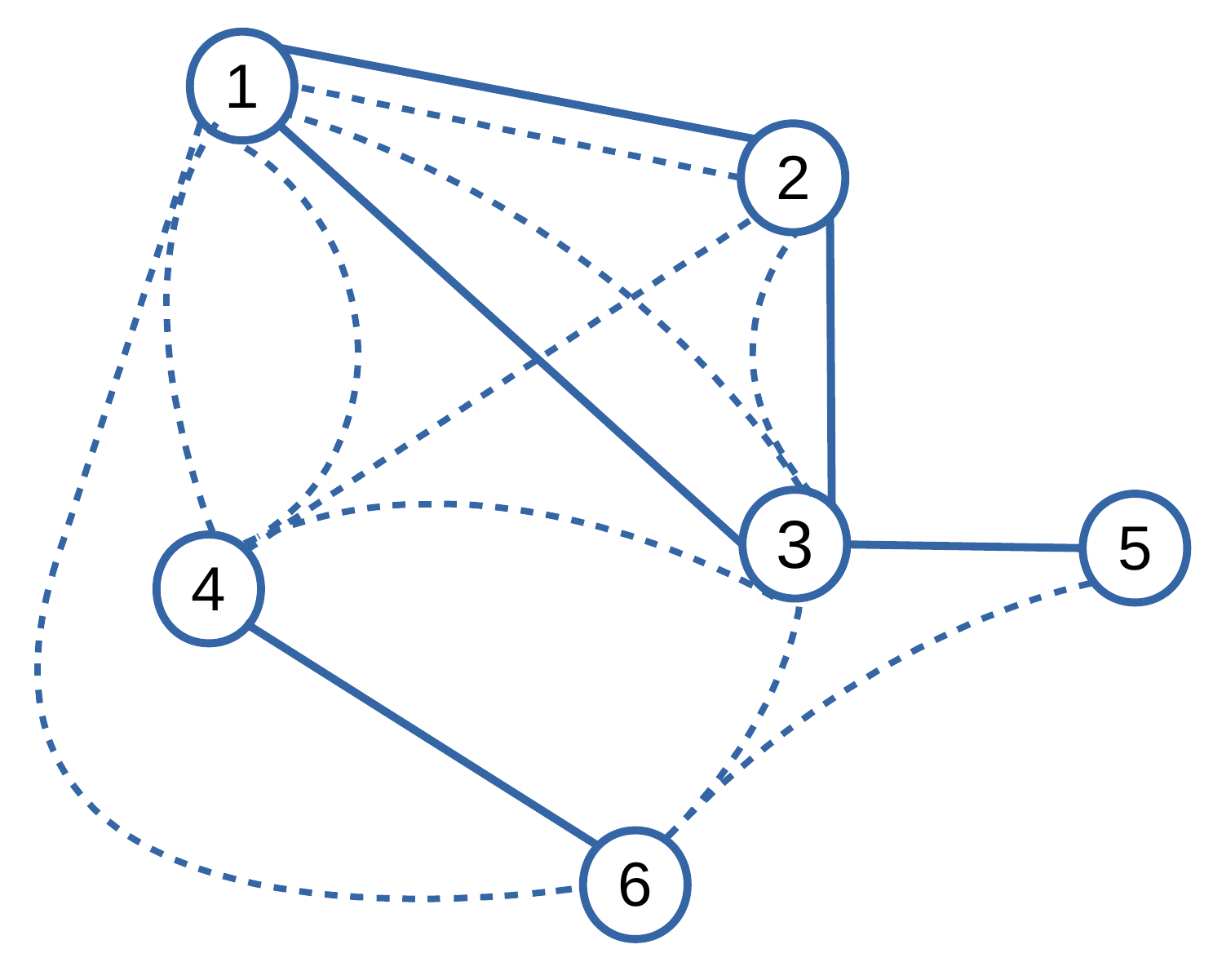}} 
  \hspace{0.2in}
  \subfigure[]{\includegraphics[width=1.2in]{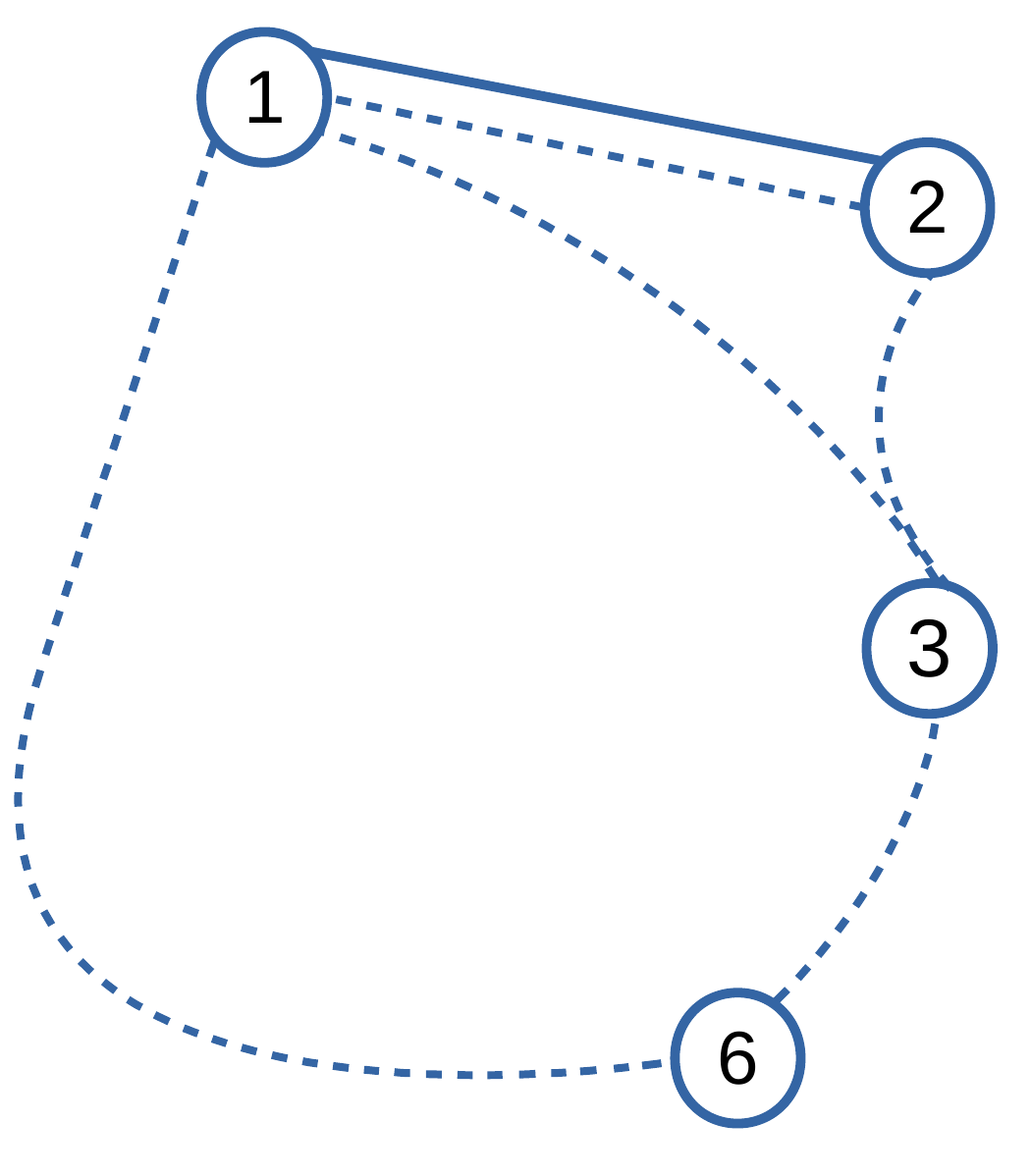}}
  \vspace{-0.1in}
  \caption{(a) Alignment graph and conflict hypergraph for an example index coding problem. (b) Alignment graph and conflict hypergraph for the restricted index coding problem, restricted to messages $1,2,3,6$. The conflict edge $\{1,2\}$ is a restricted internal conflict.}
  \label{fig:restrictedIC}
\end{figure}

The proof of the following theorem is a direct application of Theorem \ref{ratehalfthm}, and hence is skipped.
\begin{theorem}
\label{restrictedratehalfthm}
The restricted index coding problem ${\mathbb I}_{{\cal W}'}$ is rate $\frac{1}{2}$ feasible if and only if there are no ${\cal W}'$-restricted internal conflicts.
\end{theorem}
\begin{corollary}
\label{3cross1restrictedcorr}
For a given index coding problem $\mathbb I$, there exists an assignment of $3\times 1$ vectors from a two dimensional vector space (over a large enough field $\mathbb F$) to the messages in a subset ${\cal W}'$ such that the conflicts within ${\cal W}'$ are resolved, if and only if the restricted index coding problem ${\mathbb I}_{{\cal W}'}$ is rate $\frac{1}{2}$ feasible. 
\end{corollary}
\begin{IEEEproof} We first recall that conflicts within ${\cal W}'$ are said to be resolved if (\ref{resolvedconflicts}) is satisfied.

\textit{If part:} The proof for the if-part follows the achievability scheme shown in the proof of Theorem \ref{ratehalfthm}, with the difference that we assign to the messages in ${\cal W}'$ randomly generated $3\times 1$ vectors (not $2\times 1$ vectors as in Theorem \ref{ratehalfthm}) from a two dimensional space over a suitably large field $\mathbb F$. Because ${\mathbb I}_{{\cal W}'}$ is rate $\frac{1}{2}$ feasible, such an assignment resolves the conflicts within ${\cal W}'$.

\textit{Only if part:} Suppose there is an assignment of $3\times 1$ vectors from a two dimensional vector space to messages in ${\cal W}'$ such that all conflicts within ${\cal W}'$ are resolved. Then we can always obtain a $2\times 3$ matrix $A$ such that premultiplying all the vectors assigned to ${\cal W}'$ with $A$ gives us a rate $\frac{1}{2}$ (length 2) index coding solution for ${\mathbb I}_{{\cal W}'}$.
\end{IEEEproof}

The following theorem gives a necessary and sufficient condition for assigning vectors from a two dimensional space to the type-2 alignment sets (i.e., for satisfying the necessary condition of Theorem \ref{type2sets2dim}).
\begin{theorem}
\label{norestrconflictstype2}
Let ${\cal W}'$ be a type-2 alignment set of the given index coding problem $\mathbb I$. If $\mathbb I$ is rate $\frac{1}{3}$ feasible, then ${\mathbb I}_{{\cal W}'}$ must be rate $\frac{1}{2}$ feasible which holds if and only if there are no ${\cal W}'$-restricted internal conflicts.
\end{theorem}
\begin{IEEEproof}
The proof follows by combining the claims of Theorem \ref{type2sets2dim}, Corollary \ref{3cross1restrictedcorr} and Theorem \ref{restrictedratehalfthm}.
\end{IEEEproof}

\subsection{A new class of index coding problems with rate $\frac{1}{3}$ feasibility}
\label{rateonethirdfeasibleproblems}
We now prove the main result of this paper, which connects all the previously proved results and widens the class of index coding problems for which rate $\frac{1}{3}$ is achievable. Because of the framework we have developed in the previous subsections, the proof of this theorem is simpler than Theorem \ref{thmnocyclesforks}, while also subsuming that result.
\begin{theorem} \label{thm:main}
A rate $\frac{1}{2}$ infeasible index coding problem $\mathbb I$ is rate $\frac{1}{3}$ feasible if every alignment set of $\mathbb I$ satisfies\textit{ either} of the following conditions.
\begin{enumerate}
\item It does not have both forks and cycles.
\item It is a type-2 alignment set with no restricted internal conflicts.
\end{enumerate}
\end{theorem}
\begin{IEEEproof}

We first make some observations before giving the achievable index coding scheme. 

\textit{Observation 1:} Consider an alignment set ${\cal A}$ of ${\mathbb I}$ which does not contain both forks and cycles and is also not a triangular interfering set. We claim that there is no triangular interfering set within ${\cal A}$. This is because if there is a triangular interfering set (say ${\cal W}_1$) then there should be at least one more message which is not included within ${\cal W}_1$. However this would mean that there is both a fork and a cycle within ${\cal A}$, which is not allowed. 

\textit{Observation 2: }Now suppose there are three messages in ${\cal A}$ such that all of them interfere with a particular receiver. Then there must necessarily be no conflicts in-between the three messages. This is because if there were conflicts in between the three messages, then these three messages will form a triangular interfering set. Thus no three messages from ${\cal A}$ having at least one conflict in-between themselves can interfere at the same receiver.

\textit{Observation 3: }By a similar argument as in Observation $1$, if ${\cal A}$ has three messages interfering at a receiver, then it cannot have any other message than these three messages (because if it did, then we would have both cycles and forks within ${\cal A}$.

\textit{Observation 4:} Consider an alignment set ${\cal B}$ of ${\mathbb I}$ which does not contain both forks and cycles, but is also a triangular interfering set. Then, by definition ${\cal B}$ must be a type-2 alignment set. For the sake of this proof, we consider an alignment set such as $\cal B$ as being under the class of type-2 alignment sets.

We thus have three kinds of alignment sets in $\mathbb I$. 
\begin{enumerate}
\item Alignment sets which have no three messages interfering at any receiver.
\item Alignment sets which consists only of three messages, all three interfering at some receiver, without any conflicts in-between. (these three messages may interfere at other receivers also, but at least one common receiver where they all interfere exists).
\item Alignment sets which are also type-2 alignment sets without restricted internal conflicts. 
\end{enumerate}

\begin{figure}[ht]
\centering
\includegraphics[width=2.6in]{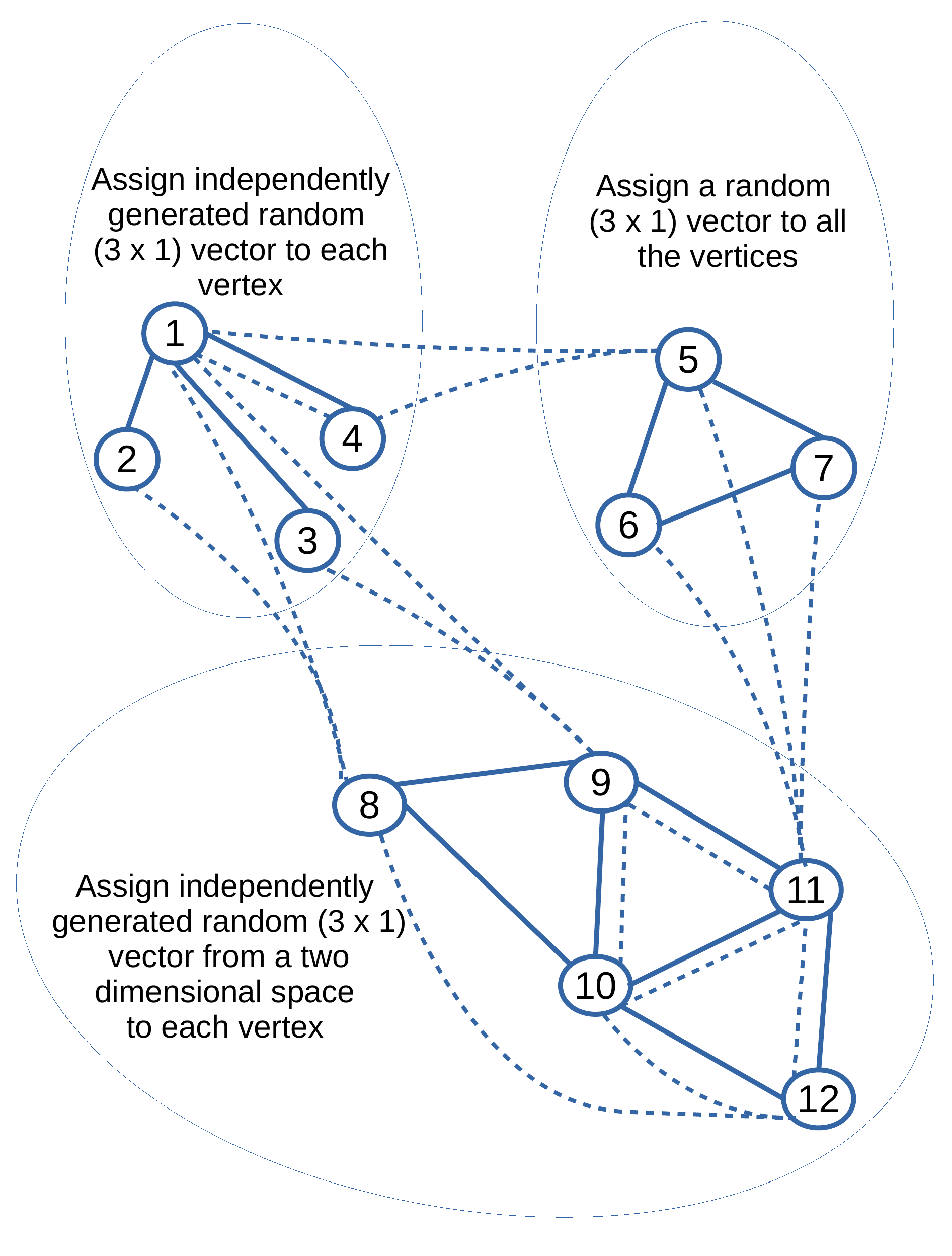}
\caption{Three types of alignment sets discussed in the proof of Theorem \ref{thm:main} are shown. In the type-2 alignment set, the receivers at which the triangular interfering sets interfere are not explicitly shown in the figure. Since there should not be any restricted internal conflict within a type-2 alignment set, no triangular interfering set interferes at a receiver inside the same type-2 alignment set.}	
\label{fig:maintheorem}	
\end{figure}

We now give the achievability index coding scheme by assigning vectors independently to each alignment set of $\mathbb I$, and the technique of assignment, which follows, depends on the type of the alignment set. 

\textit{Alignment set which has no three messages interfering at any receiver:} For each message in such an alignment set, we assign an independently generated random $3\times 1$ vector (over a large field $\mathbb F$).

\textit{Alignment set which consists only of three messages interfering at any receiver without any conflicts in-between:}
We randomly generate a $3 \times 1$ vector and assign it to all the messages in such an alignment set.

\textit{Alignment set which is a type-2 alignment set without restricted internal conflicts:}
Let ${\cal W}'$ be the type-2 alignment set under concern. For each ${\cal W}'$-restricted alignment set, we assign an independently generated random $3\times 1$ vector from a two dimensional space. Note that this resolves all the conflicts between the messages within ${\cal W}'$, by Theorem \ref{restrictedratehalfthm} and Corollary \ref{3cross1restrictedcorr}. 

All the messages fall under one of these alignment sets, and hence all of the messages have been assigned vectors at this point. Let $\mathbb E$ denote encoding function corresponding to this assignment and $V_k$ denote the vector assigned to message $W_k$. We now show that this assignment resolves all the conflicts in $\mathbb I$.

Consider a receiver $j$ which requests a message $W_k$. We proceed on a case by case basis, depending on the size of $Interf_k(j)$. For each case we check whether the condition, $V_k\notin V_{\mathbb E}(Interf_k(j))$ (whp), is met. We call this condition as the \textit{no conflict condition} for the sake of this proof.

\textit{Case 1:} $|Interf_k(j)|\leq 2:$
There are two cases here, either $W_k$ and $Interf_k(j)$ are in the same alignment set or they are in different alignment sets. If $W_k$ and $Interf_k(j)$ are in different alignment sets, then $W_k$ and the messages $Interf_k(j)$ have been assigned independently and randomly generated $3\times 1$ vectors. Thus the no conflict condition is met.

Now,  $W_k$ and $Interf_k(j)$ are in the same alignment set. This can only be an alignment set where no three messages interfere at any receiver, or a type-2 alignment set. In the former case, the no conflict condition is met as any three messages in such an alignment set are assigned independent vectors (whp). Now if it is a type-2 alignment set, the conflict(s) between $W_k$ and $Interf_k(j)$ is(are) within that type-2 alignment set. Because there are no restricted internal conflicts in any type-2 alignment set, it must be the case that $W_k$ and $Interf_k(j)$ are in different restricted alignment sets. By our scheme, such conflicts should therefore be resolved. Hence, the no conflict condition is met in this case too.

\textit{Case 2:} $|Interf_k(j)|\geq 3:$ Then we have two cases. The first case is that no two messages in $Interf_k(j)$ are in conflict. This means that $Interf_k(j)$ must be an alignment set which consists only of three messages without any conflicts in-between, and $W_k$ necessarily belongs to another alignment set. By our scheme, $V_{\mathbb E}(Interf_k(j))$ is an one-dimensional space generated by a random vector which is independently generated from the vector assigned to $W_k$. Hence, the no conflict condition is met.

Finally we consider the case when $|Interf_k(j)|\geq 3$ and at least one conflict exists within $Interf_k(j)$. Then $Interf_k(j)$ must be within some type-2 alignment set (say ${\cal W}'$). Again we have two sub-cases here, i.e., $W_k$ and $Interf_k(j)$ are within the same (type-2) alignment set, or $W_k$ is in a different alignment set than $Interf_k(j)$ which is within a type-2 alignment set. In the former case, the conflicts between $W_k$ and $Interf_k(j)$ are within the type-2 alignment scheme, which is resolved by our scheme (by the same arguments as in the last subcase of Case 1). In the latter case, we must have that $V_{\mathbb E}(Interf_k(j))$ lies within a two dimensional space (as it is within a type-2 alignment set) which is generated independently from the vector assigned to $W_k$. Hence again the no conflict condition is met.

By the previous arguments, the no conflict condition is met for any receiver $j$ and any demand $W_k$ at $j$. Thus, all the conflicts in $\mathbb I$ are resolved. This proves the theorem.
\end{IEEEproof}
\section{Discussion}
In this work, we presented a class of index coding problems for which rate $\frac{1}{3}$ is feasible. This class of problems is larger than what was previously known. We believe that the framework developed in this work in order to obtain our results can be leveraged to settle the rate $\frac{1}{3}$ feasibility completely. In particular, we conjecture that the the necessary condition for rate $\frac{1}{3}$ feasibility of Theorem \ref{norestrconflictstype2} is also sufficient.
\begin{conjecture*}
A given index coding problem is rate $\frac{1}{3}$ feasible if and only if all the type-2 alignment sets have no restricted internal conflicts.
\end{conjecture*}
Developing conditions for feasibility of rates of the form $\frac{1}{m}$ would also be an interesting area of future study. The connection to topological interference management problems follows from \cite{Jaf}. From \cite{BBJK}, it is known that the length of scalar linear index codes for the single unicast index coding problem is known to be equal to \textit{minrank} of the side-information graph. Thus, our results also imply a class of graphs whose minrank is equal to $3$. This is a promising result as the general minrank problem is known to be NP hard \cite{Pee}.


\begin{thebibliography}{160}
\bibitem{BiK} Y. Birk and T. Kol, ``Coding on Demand by an Informed Source (ISCOD) for Efficient Broadcast of Different Supplemental Data to Caching Clients``, IEEE Transactions on Information Theory, Vol. 52, No. 6, June, 2006, pp. 2825-2830.

\bibitem{BBJK}
Z.Bar-Yossef, Y. Birk, T.S. Jayram, T. Kol, ``Index Coding with Side Information'', IEEE Transactions on Information Theory, Vol. 57, No. 3, March 2011, pp. 1479-1494.

\bibitem{SDL}
K. Shanmugam, A. G. Dimakis, and M. Langberg, ``Local graph coloring and index coding'', ArXiv, Feb. 2013, Available at http://arxiv.org/abs/1301.5359.

\bibitem{BKL1}
A. Blasiak, R. Kleinberg, and E. Lubetzky, ``Index Coding via Linear Programming'', ArXiv-CoRR, Jul. 2011, Available at http://arxiv.org/abs/1004.1379.

\bibitem{BKL2}
 A. Blasiak, R. Kleinberg, and E. Lubetzky, ''Broadcasting with side information: Bounding and approximating the broadcast rate'', IEEE Transactions on Information Theory, Vol. 59, No. 9, Sep. 2013, pp. 5811-5823.


\bibitem{ArK}
F. Arbabjolfaei, Y-H Kim, ``Structural properties of index coding capacity using fractional graph theory'', IEEE ISIT 2015, Hong Kong, 14-19 June, pp.1034-1038.

\bibitem{TOJ}
C. Thapa, L. Ong, S. J. Johnson, ``Generalized interlinked cycle cover for index coding'', IEEE ITW (Fall) 2015, Jeju Island, South Korea, Oct. 11-15 2015.

\bibitem{LuS}
E. Lubetzky and U. Stav, ``Nonlinear Index Coding Outperforming the Linear Optimum'', IEEE Transactions on Information Theory,  Vol. 55, No. 8, Aug. 2009, pp. 529-568.


\bibitem{ABKSW}
F. Arbabjolfaei, B. Bandemer, Y-H Kim, E. Sasoglu, L. Wang, ``On the capacity region for index coding'', IEEE ISIT 2013, Istanbul, Turkey, 7-12 July, pp. 962-966.

\bibitem{TDN}
 A. S. Tehrani, A. G. Dimakis, and M. J. Neely, ``Bipartite index coding'', IEEE ISIT 2012, Cambridge, MA, USA, Jul 1-6, pp. 2246-2250.


\bibitem{MCJ1}
H. Maleki, V. Cadambe, and S. Jafar, ``Index coding: An interference alignment perspective'', IEEE ISIT 2012, Cambridge MA, USA, Jul 1-6, pp. 2236-2240.

\bibitem{Jaf} 
S. A. Jafar, ``Topological Interference Management Through Index Coding'', IEEE Transactions on Information Theory, Vol. 60, No. 1, Jan. 2014, pp. 529-568.


\bibitem{MCJ2}
H. Maleki, V. R. Cadambe, and S. A. Jafar, ``Index Coding — An Interference Alignment Perspective'', IEEE Transactions on Information Theory, Vol. 60, No. 9, Sep. 2014, pp. 5402-5432.




\bibitem{Pee}
 R. Peeters, ``Orthogonal representations over finite fields and the chromatic number of graphs'', Combinatorica, Vol. 16, No. 3, pp. 417–431, Sept 1996.
\end{thebibliography}
\end{document}